\documentclass[twocolumn]{aastex62}

\usepackage{soul, xcolor}
\setstcolor{red}
\usepackage{multirow}

\usepackage{amsmath}

\newcommand{\hi}{\textsc{Hi}}
\newcommand{\cmmc}{\textsc{21cmMC}}

\usepackage[caption=false]{subfig}

\graphicspath{{./}{figures/}}

\received{January 1, 2018}
\revised{January 7, 2018}
\accepted{\today}
\submitjournal{ApJ}

\shorttitle{Impacts of Foreground and Instrument on EoR Signal}
\shortauthors{Nasirudin et al.}

\begin{document}

\title{The Impact of Realistic Foreground and Instrument Models on 21\,cm Epoch of Reionization Experiments}

\correspondingauthor{A.~ Nasirudin}
\email{bella.nasirudin@icrar.org}

\author[0000-0002-0786-7307]{A.~Nasirudin}
\affil{International Centre for Radio Astronomy Research - Curtin University, Bentley, WA 6102, Australia}
\affiliation{ARC Centre of Excellence for All Sky Astrophysics in 3 Dimensions (ASTRO 3D)}

\author{S.~G.~Murray}
\affil{International Centre for Radio Astronomy Research - Curtin University, Bentley, WA 6102, Australia}
\affiliation{ARC Centre of Excellence for All Sky Astrophysics in 3 Dimensions (ASTRO 3D)}
\affiliation{Arizona State University, Tempe, AZ, USA}

\author{C.~M.~Trott}
\affil{International Centre for Radio Astronomy Research - Curtin University, Bentley, WA 6102, Australia}
\affiliation{ARC Centre of Excellence for All Sky Astrophysics in 3 Dimensions (ASTRO 3D)}

\author{B.~Greig}
\affiliation{ARC Centre of Excellence for All Sky Astrophysics in 3 Dimensions (ASTRO 3D)}
\affiliation{School of Physics - The University of Melbourne, VIC 3010, Australia}

\author{R.~C.~Joseph}
\affil{International Centre for Radio Astronomy Research - Curtin University, Bentley, WA 6102, Australia}
\affiliation{ARC Centre of Excellence for All Sky Astrophysics in 3 Dimensions (ASTRO 3D)}

\author{C.~Power}
\affiliation{ARC Centre of Excellence for All Sky Astrophysics in 3 Dimensions (ASTRO 3D)}
\affiliation{International Centre for Radio Astronomy Research - The University of Western Australia, Crawley, WA 6009, Australia}


\begin{abstract}

Predictions for the ability of 21-cm interferometric experiments to discriminate Epoch of Reionization (EoR) signal models are typically limited by the simplicity of data models, whereby foreground signals and characteristics of the instrument are often simplified or neglected.
To move towards more realistic scenarios, we explore the effects of applying more realistic foreground and instrument models to the 21cm signal, and the ability to  estimate astrophysical parameters with these additional complexities.
We use a highly-optimized version of \textsc{21cmFAST}, integrated into \textsc{21cmMC}, to generate lightcones of the brightness temperature fluctuation for Bayesian parameter estimation. We include a statistical point-source foreground model and an instrument model based on the Murchison Widefield Array (MWA) scaled in observation time to have an effective sensitivity similar to the future Square Kilometre Array (SKA). We also extend the current likelihood prescription to account for the presence of beam convolution and foregrounds, the 2-Dimensional Power Spectrum (PS), and the correlation of PS modes. We use frequency bands between 150 and 180 MHz to constrain the ionizing efficiency ($\zeta$), the minimum virial temperature of halos ($T_{\mathrm{vir}}$), the soft X-ray emissivity per unit Star Formation Rate (SFR) ($L_X/SFR$ ), and the X-ray energy threshold ($E_0$). We find that the inclusion of realistic foregrounds and instrumental components biases the parameter constraints due to unaccounted for cross-power between the EoR signal, foregrounds and thermal noise. This causes estimates of $\zeta$ to be biased by up to $5\sigma$ but the estimates of $T_{vir}$, L$_X$/SFR and E$_0$ remain unaffected and are all within $1\sigma$. 

\end{abstract}


\section{Introduction} \label{sec:intro}

The Cosmic Dawn (CD) and the subsequent Epoch of Reionization (EoR) mark the end of the cosmic dark ages, during which time the baryonic content of the dark, early universe existed in a warm, neutral state. The intergalactic medium (IGM), predominantly comprised of hydrogen, was illuminated by photons from the first ionizing sources, forming regions of reionized hydrogen with temperatures contrasting with the neutral IGM and background CMB (\citealt{barkana2001beginning}; \citealt{furlanetto2006cosmology}). The significance of these periods is prevalent in almost all areas of astrophysics, particularly in understanding the transition between the current and early universe as well as the formation of primordial structures in the universe.

A key probe into these epochs is the imprint of the 21\,cm spin-flip transition of neutral hydrogen (\hi) that is redshifted into the low-frequency radio regime. 
Recently, the Experiment to Detect the Global Epoch of Reionization Signature (EDGES) published a measurement of a flattened absorption profile with timing that somewhat coincides with the expected \hi\ signal from the CD \citep{bowman18}, although it needs further verification from independent experiments. Detection of the EoR, however, has not been reported and remains one of the key science goals of most current low-frequency interferometric telescopes (e.g., the Murchison Widefield Array (MWA) (\citep{tingay2013murchison,wayth2018}), the Hydrogen Epoch of Reionization Experiment (HERA) \citep{deboer2017hydrogen}, and the Low Frequency Array (LOFAR) \citep{van2013lofar}).

Since direct imaging of the EoR remains beyond the sensitivity capabilities of current experiments, these instruments aim to detect the spatial fluctuations of the averaged power spectrum. In the near future, however, upcoming low-frequency interferometers such as the Square Kilometre Array (SKA) \citep{dewdney2009square} and the final phase of HERA are expected to have the sensitivity to directly detect the tomographic imprints of the EoR.

Even with experiments focused on the detection of the variance instead of direct imaging of the EoR signal, its detection is still challenging due to astrophysical foregrounds and other contaminants. These include foregrounds from Galactic and extragalactic origins (\citealt{jelic2010realistic}; \citealt{gleser2008decontamination}),  ionospheric distortion \citep{jordan2017characterization}, instrument noise and radio frequency interference \citep{bentum2008implementation, offringa2015low}. The foregrounds prove to be the primary contaminant as they are expected to be up to 5 orders of magnitude brighter than the 21\,cm signal \citep{pritchard201221}. Various foreground mitigation and removal methods have been extensively studied (e.g. see \cite{liu2019data} for details); however,  for foreground removal method such as source peeling, residual power is still left in the power spectrum due to our incomplete knowledge of the extra-galactic foregrounds. This is due to imperfect peeling and fainter sources existing below the peeling threshold \citep{datta10, trott2012impact,morales2012four,vedantham12}\footnote{In this work, we assume that the former is negligible and focus on the latter.}.

Exploration of a physical understanding of the EoR is performed through theoretical simulations of the sources and signal to help supplement observations and constrain the signal. In particular, most simulation work has focused on establishing the roles of the first ionising sources and finding the impact of astrophysical parameters on the \hi\ signal (see eg. \citealt{barkana2001beginning}; \citealt{ricotti2004x}; \citealt{madau2004early}; \citealt{knevitt2014heating}). Although countless parameters may affect the EoR signal whose estimates vary across the literature, some of the most influential EoR parameters (and their generally-accepted estimates) are: the number of ionizing photons per baryon (N$_\gamma\sim$ 4000) \citep{barkana2005method}; the power-law scale of the baryonic gas fraction in stars with source halo mass ($\alpha_* \approx $ 0.5) \citep{behroozi2015simple,ocvirk2016cosmic}; and the number of times a hydrogen atom recombines (n$_{rec} \approx $ 1) \citep{sobacchi2014inhomogeneous}.

The ultimate goal of understanding the EoR from 21\,cm observations will require constraining these parameters using theoretical models . It involves quantitatively evaluating the cosmic EoR signal by making use of one of the many existing EoR simulations, usually semi-numerical, to extract information on the astrophysics. Aside from comparing the results of fiducial reionization simulations with experimental data \citep{choudhury2005experimental}, other methods used to achieve this goal include using the maximum likelihood ($\chi^2$) fitting \citep{barkana2009studying}, and Bayesian analyses via model selection \citep{binnie2019bayesian} or Monte Carlo Markov Chain (MCMC) methods \citep[see eg.][]{harker2011mcmc,patil2014constraining,greig201521cmmc,hassan2017epoch}\footnote{The recent work of \cite{sims2019joint} also uses an MCMC framework, but focuses on estimating the power spectrum band-powers rather than astrophysical parameters, though it mentions a simple extension to do so.}. Alternative methods to on-the-fly MCMC sampling include emulating simulations using the power spectrum \citep{kern2017emulating} and using artificial neural networks \citep{shimabukuro17,schmit2017neural} or convolutional neural network \citep{gillet2019deep}.

A particularly powerful existing code that uses the MCMC approach is \cmmc\ \citep{greig201521cmmc,greig17,greig201821cmmc}. \cmmc\ is a parallelized, efficient EoR analysis code that wraps the publicly-available EoR semi-numerical simulator \textsc{21cmFAST} \citep{mesinger201121cmfast} into its Bayesian MCMC framework to produce an EoR lightcone per-iteration\footnote{In this work, we use the most recent version available at \url{https://github.com/BradGreig/Hybrid21cm}. As of writing, the most up-to-date version of \cmmc\ has moved permanently to \url{https://github.com/21cmFAST/21cmMC}, and the underlying optimized \textsc{21cmFAST} code has been modularized and exists at \url{https://github.com/21cmFAST/21cmFAST}.}. It uses the differential brightness temperature field, constructed as a lightcone, to compute the spherically averaged power spectrum that is, in turn, used in the likelihood prescription. 

Previous applications of \cmmc\ have used \textsc{21cmSense}\footnote{https://github.com/jpober/21cmSense} to gauge the uncertainties in the 1-Dimensional (1-D) power spectrum arising from the instrumental noise and smooth foregrounds, completely excising spherical $k$-modes dominated by the latter \citep{pober2013baryon,pober2014next}.  While this approach is a good first-order approximation, it does not account for potential residual foreground contamination in $k$-modes that it uses due to more complicated instrumental systematics. 
This can cause difficulties in disentangling contamination and signal power, thus possibly leading to a biased estimate of reionization parameters. Moreover, it does not account for the induced chromaticity from the instrument on the 21\,cm signal.

This work provides an extension to \cmmc,  in the publicly-available plug-in \textsc{py21cmmc-fg}. Using this framework, we aim to explore the impact of foreground and instrumental components on the ability to constrain the astrophysical parameters of the EoR. We apply a statistical point-source foreground model and an MWA-based instrument model to the \cmmc\ lightcone outputs before averaging to a power spectrum. It is these ``corrupted'' lightcones that we use to estimate observed power spectra within the MCMC framework, thereby investigating more realistic constraints on astrophysical parameters.

The paper is structured as follows. We first setup the mathematical framework, which includes describing the foreground and instrumental effects we employ in our analysis, in \S \ref{sec:methodology}. In \S \ref{sec:EoR_analysis}, we analyze the observational effects on the EoR lightcones and in \S \ref{sec:mcmc_analysis}, we present the MCMC analysis using \cmmc\ with our pipeline. We then discuss the results in \S \ref{sec:results} and conclude in \S \ref{sec:discussion}.

\section{Mathematical Framework} \label{sec:methodology} 

\subsection{Interferometric Visibilities}
The baseline displacement, \textbf{u}, is defined as $\mathbf{u} = \mathbf{x}/\lambda$, where \textbf{x} is the physical displacement between the baseline tiles and $\lambda$ is the wavelength. We define the sky coordinate as $\mathbf{l} = (l, m) = (\sin \theta \cos \phi, \sin \theta \sin \phi$), where $\theta$ is the zenith angle, and $\phi$ the angle around the zenith pole.

Based on these definitions, the measured correlation of the electric fields between two sensors for an interferometric observation, the visibility $V(\textbf{u},\nu)$ at frequency $\nu$, in the flat-sky approximation is defined as
 \begin{equation}
 \label{eqn:FT_foreground}
 V(\textbf{u},\nu) = \int I(\textbf{l}, \nu) B (\textbf{l}, \nu) \exp(-2\pi i \textbf{u}\cdot \textbf{l}) d\textbf{l} \ \ \ \mathrm{ [Jy]},
 \end{equation}
 with  $I(\textbf{l}, \nu)$ and $B(\textbf{l}, \nu)$ being the intensity of each point-source and beam attenuation at sky coordinate \textbf{l} and frequency $\nu$, respectively. Using the flat-sky approximation whereby the effects of curvature are neglected, we assume that the observed interferometric visibility is identical to the Fourier Transform of the product of signal and the beam model. We will address the second-order effects of spatial curvature on the visibility in future work. 

\subsection{Brightness Temperature and Power Spectrum}
\label{sec:power_spectrum}
The EoR differential brightness temperature, $\delta T_{b}$, can be quantified by \citep{furlanetto2006cosmology}
\begin{align}
\label{eqn:Tb}
\delta T_{b} (z) \approx & 27 x_{HI} (1+\delta_{nl}) \left(\frac{H(z)}{\mathrm{d}v/\mathrm{d}r +H(z)}\right)\left(1 -\frac{T_\gamma}{T_s}\right) \nonumber \\
& \times \left(\frac{1+z}{10}\frac{0.15}{\Omega_m h^2}\right)^{\frac{1}{2}}\left(\frac{\Omega_b h^2}{0.023}\right)\ [{\rm mK}  ].
\end{align}
Here, $x_\mathrm{HI}$ is the neutral fraction, $\delta_\mathrm{nl}$ is the evolved Eulerian overdensity, $H$ is the evolving Hubble constant, d$v$/d$r$ is the gradient of the line-of-sight velocity component, $T_\gamma$ is the temperature of the CMB, $T_\mathrm{s}$ is the spin temperature of \hi, $z$ is the redshift, $\Omega_m$ is the dimensionless matter density parameter, $\Omega_b$ is the dimensionless baryonic density parameter and $h$ is the normalized Hubble constant.

For interferometric observations, the power spectrum (hereafter PS), is the primary metric used to characterise the EoR signal. It measures the spatial variance of a signal over a spatial volume $V$ and is defined as
\begin{equation}
\label{eqn:ps}
P(k) \equiv \frac{|  \langle \delta T_b^\dagger(\vec{k}) \delta T_b(\vec{k})\rangle_{|\vec{k}|= k}|}{V}\ \ \  [ {\textrm{mK}}^2 { \textrm{ Mpc}^{-3} h^3}],
\end{equation}
where $k$ is the spatial scale in Fourier space. The PS can be computed either directly from the image cube or from the observed interferometric visibilities, whereby the signal is spherically-averaged over the 3-D spatial scales and is normalised by the volume of the observed area of the sky. The dimensionless 1-D PS is given by
\begin{equation}
\label{eqn:ps_dimensionless}
\Delta^2(k) = \frac{k^3}{2 \pi^2} P(k)\ \ \ [{\rm mK}^2 ],
\end{equation}
and is routinely used in current experiments.

It is useful, however, to first compute the cylindrically-averaged 2-D PS,  $P(k_{\bot},k_{\parallel})$. The angular and line-of-sight modes of the 2-D PS, $k_\perp$ and $k_\parallel$, are converted from the Fourier dimensions following \citep{morales2010reionization}
\begin{equation}
\label{eqn:kperp_modes}
k_\perp =  \frac{2 \pi |\textbf{u}|}{D_M(z)}\ \ \  [ {\rm Mpc^{-1} h}],
\end{equation}
and
\begin{equation}
\label{eqn:kpar_modes}
k_\parallel =  \frac{2 \pi H_0 f_{21} E(z) }{c(1+z)^2} \eta \ \ \  [ {\rm Mpc^{-1} h}].
\end{equation}
Here, $z$ is the observation redshift, $D_M(z)$ is the transverse comoving distance, $H_0$ is the Hubble constant, $f_{21}$ is the rest frequency of the 21\,cm hydrogen hyperfine transition and $E(z)$ is defined as 
 \begin{equation}
\label{eqn:Ez_cosmology}
E(z) =  \sqrt[]{\Omega_m(1+z)^3+\Omega_k(1+z)^2+\Omega_\Lambda},
\end{equation}
where $\Omega_\Lambda$, and $\Omega_k$ are the dimensionless density parameters for dark energy and the curvature of space \citep{hogg1999distance}. Because the 2-D PS bins the signal into separable perpendicular and line-of-sight modes, it effectively allows for the management of systematic effects arising from the different modes, and hence is commonly used as an initial step to decontaminate the cosmological EoR signal from the foregrounds. Colloquially termed the ``wedge", the broad region of cylindrical $k$-space dominated by the foregrounds can be understood as the signature of smooth-spectrum foregrounds when being sampled by an imperfect instrument \citep{vedantham12,trott2012impact,parsons12,datta10}. The key imperfections are the band-limiting attenuation due to the primary beam, and the discrete sampling of an interferometer, where the sampled wavemodes change with frequency (``chromaticity").

\subsection{Components of Observable Signal}
\label{sec:components_all}
In this subsection, we present the mathematical framework of the different components of the observable signal, which is made of the EoR signal and foregrounds as observed by our instrument model.

\subsubsection{EoR Lightcones and Parameters }
\label{sec:signal_component}
We use a model $\delta T_b$ field generated by \textsc{21cmFAST} wrapped in \cmmc.
\textsc{21cmFAST} is a semi-numerical EoR modelling tool designed to efficiently simulate the 21-cm \hi\ signal using approximate methods combining the excursion-set formalism \citep{bond1991excursion,furlanetto2004growth} and perturbation theory. The code generates realizations of $\delta_{nl}$, ionization, peculiar velocity, and $T_s$ in 3D which are then combined to compute $\delta T_b$ during the EoR. The astrophysical parameters involved in the code are customizable, allowing for the exploration of models and parameter space affecting the EoR \citep{mesinger201121cmfast,mesinger14}.

The key parameters that we are primarily interested in are:
\begin{itemize}
\item the ionizing efficiency i.e., the number of ionizing photons escaping into the IGM per baryon ($\zeta$ [dimensionless])
\item the minimum virial temperature of halos required to form stars in galaxies ($T_{\mathrm{vir}}$ [K])
\item the soft X-ray emissivity per unit Star Formation Rate (SFR) escaping galaxies ($L_X/SFR$ [erg s$^{-1}$ keV$^{-1}$ M$^{-1}_\odot$ yr])
\item the X-ray energy threshold for self-absorption by galaxies ($E_0$ [keV]).
\end{itemize}

These four parameters are chosen due to their relatively high influence on the $\delta T_b$ field, as they directly affect the parameters governing $T_b$ given by Eq.~\ref{eqn:Tb}, particularly $x_\mathrm{HI}$\footnote{\cmmc\ now includes an updated parameterization that uses $>$6 parameters, but we chose this ``legacy" set for ease of comparison to previous work.}. The fiducial values (and ranges) we adopt are based on \cite{greig17}, \cite{park2019inferring} and \cite{gillet2019deep} and summarised in Table~\ref{tbl:params}. We note that we only use these values in our final analysis in \S \ref{sec:results}. For exploration and validation purposes presented in \S \ref{sec:EoR_analysis} and \ref{sec:mcmc_analysis}, we use the default parameters of $\zeta = 30$ and log$_{10}(T_{vir}/\rm K) = 4.7$. We also use the same random seed to generate the signal throughout this research.

\begin{table}[]
\centering
\begin{tabular}{|l|l|l|}
\hline
\textbf{Parameter} & \textbf{Fiducial Value} & \textbf{Range}       \\ \hline
$\zeta$            & 20.0                    & {[}10, 250{]}        \\ \hline
log$_{10} (T_{\rm vir}$)            & 4.48         & {[}4, 6{]} \\ \hline
log$_{10} (L_X/$SFR)          & 40.5                    & {[}38, 42{]}         \\ \hline
E$_0$              & 0.5                     & {[}0.1, 1.5{]}       \\ \hline
\end{tabular}
\caption{Fiducial astrophysical parameters and their ranges adopted in this work.}
\label{tbl:params}
\end{table}

We assume a $\Lambda$ Cold Dark Matter ($\Lambda$CDM) universe using the default cosmological parameters values of \textsc{21cmFAST} ($h$ = 0.68, $\Omega_b$=0.048, $\Omega_m$= 0.31, $\Omega_k$=0, and $\Omega_\Lambda$= 0.69), consistent with results from \cite{planck2016ade}. 

The conversion from $T_b$ to flux density $S(\nu)$ is given by the beam-modified Rayleigh-Jeans law,
\begin{equation}
\label{eqn:mK_to_Jy}
S(\nu) = \left(\frac{2k_BT_b}{A_{\rm eff}}\right) \Omega \times 10^{26}  \ \ \ [{\rm Jy}],
\end{equation}
where $k_B$ is the Boltzmann constant, $A_{\rm eff}$ is the effective area of the tile (units of m$^2$) and $\Omega$ is the angular size of the beam (units sr).

\subsubsection{Point-Source Foreground Model}
\label{sec:foreground_component}
We use a simple point-source foreground model to simulate the effects of extra-galactic foreground sources based on the power-law relation,
\begin{equation}
\label{eqn:dn_ds}
\frac{dN}{dS} (S, \nu)= \alpha S_{\nu}^{-\beta} \left( \frac{\nu}{\nu_0}\right)^{-\gamma \beta} [{\rm Jy}^{-1} {\rm sr}^{-1}].
\end{equation}

Here, $dN/dS$ is the source spatial density per unit flux density, $S_{\nu}$ is the flux at a specific frequency $\nu$, $\beta$ is the slope of the source-count function, and $\gamma$ is the mean spectral-index of point sources. 
We fiducially use $\alpha$ = 4100 Jy$^{-1}$ sr$^{-1}$,  $\beta$ = 1.59, and $\gamma = 0.8$ at $\nu_0$ = 150 MHz based on an observational result from \citet{intema2011deep}.

Our adopted statistical foreground model, while an improvement over previous modeling in the context of parameter estimation, is by no means complete. It ignores potential point-source clustering \citep{murray2017improved}, ionospheric effects \citep{jordan2017characterization,mevius16,trott18}, and more subtly, any potential correlations of the foregrounds with their antecedent EoR counterparts. More importantly, we have not included a galactic diffuse foreground model in this work. These effects are expected to be second-order, except for the diffuse emission, which is bright and spatially-structured. These components are left as extensions to this work.

\subsubsection{MWA and SKA-based Instrument Model}
\label{sec:instrument_component}
Our instrument components are based on the MWA \citep{tingay2013murchison,wayth2018} and the future SKA \citep{dewdney2009square}. The MWA is a low-frequency radio aperture array telescope located at the Murchison Radio Astronomy Observatory (MRO) site in Western Australia. The array consists of 128 connected tiles with an effective area per tile ($A_{\rm eff}$) of 21\,m$^2$ at 150\,MHz. Each tile consists of a 4x4 grid of dual polarization dipoles with a full-width half-maximum field of view of 26$^{\circ}$ at $\nu_0 = 150$ MHz. It operates in the 80 -- 300\,MHz frequency range, making it an excellent probe of the redshifted EoR signal. As one of its primary scientific goals, the MWA reionization observing scheme spans two 30\,MHz bands, between 137 -- 167\,MHz and 167 -- 197\,MHz  \citep{jacobs2016murchison}.

The low-frequency part of the future SKA (SKA-low) will be located at the MRO alongside the MWA \citep{dewdney2009square,mellema13}. It is expected to have a frequency resolution of 1\,kHz with a frequency band of 50 to 200\,MHz, and a Field-Of-View (FOV) of 2.5$^{\circ}$ -- 10$^{\circ}$.

The instruments themselves are hugely complicated; e.g., primary beam responses change between antennas, pointing and polarisation, the dipole array structure yields complex, frequency-dependent beam patterns, signal transport over coaxial cable can lead to cable reflections (imprinting frequency structure into the signal chain), and the large FOV introduces wide-field effects. We restrict ourselves to the primary instrumental response, including baseline sampling, FOV, and frequency-dependent primary beams, because these are the leading-order effects, and leave other instrumental effects to future work. We also neglect the fact that the earth is rotating and assume a fixed zenith pointing at the same patch of the sky. Neglecting the rotation of the Earth changes the $uv$ coverage of the visibilities, hence the sample variance is different, and, due to the wide FOV of the MWA, amounts to a reduction of the overall thermal noise by (only) a factor of up to three on long baselines for the same total integration time. While this should be kept in mind, it is tangential to the point of our present work, and will be explored in more detail in future work.

We approximate the beam attenuation, $B(\textbf{l}, \nu)$, to be Gaussian-shaped with
\begin{equation}
\label{eqn:beam}
B(\textbf{l}, \nu) = \exp\left(\frac{-|\textbf{l}|^2}{\sigma_{\rm beam}^2(\nu)}\right),
\end{equation}
where
\begin{equation}
\label{eqn:sigma_beam}
\sigma_{\rm beam}(\nu) \simeq \frac{\epsilon c}{\nu D} .
\end{equation}
Here, $\epsilon \simeq 0.42$  is the scaling from the more natural Airy disk to a Gaussian width, $c$ is the speed of light and $D$ is the tile diameter (4 m for the MWA). 
Although the wide field-of-view of the MWA renders the flat sky-approximation (and hence Eq.~\ref{eqn:FT_foreground}) inaccurate \citep{thyagarajan2015confirmation,thyagarajan2015foregrounds}, we will still use it as a reasonable first approximation as the curved-sky treatment will be much more important when using a more realistic beam with side-lobes, which enhances the ``pitchfork'' effect. We assume a fixed zenith pointing of the instrument over $\textbf{l} \in (-1, 1)$ and pad the sky with zeros over 3 times the size of the sky to increase the resolution of the discrete Fourier Transform.

We add thermal noise to our framework corresponding to both measurement and radiometric noise. This is the uncertainty of the visibility arising from the finite number of samples, given by 
\begin{equation}
\label{eqn:thermal_noise}
\sigma_N = 10^{26}\frac{2k_BT_{\rm sys}}{A_{\rm eff}} \frac{1}{\sqrt[]{\Delta \nu \Delta t}} \ \ \ \mathrm{ [Jy]},
\end{equation}
where $T_{\rm sys}$ is the system temperature, $\Delta \nu $ is the bandwidth of one frequency channel, and $\Delta t $ is the integration time of the observation in seconds. $\sigma_N$ is essentially an estimation of the global sky signal (or temperature) for a given set of information that is dependent on the sky temperature, bandwidth and sampling time for each visibility.

In addition, we use the unnormalized\footnote{The normalization cancels out in Equation 14.} Fourier Gaussian beam kernel in re-gridding the visibilities after baseline sampling where the weight in cell $i$, $w_i$ is given by
\begin{equation}
\label{eqn:weights}
w_i =  \exp (- [\pi \sigma_{beam}(\nu) |\textbf{u}_i-\textbf{u}_j|]^2),
\end{equation}
for baseline $j$. The gridded visibility $V_{\rm grid}(\textbf{u}, \nu)$ is, hence, given by
\begin{equation}
\label{eqn:grid_vis}
V_{\rm grid}(\textbf{u}, \nu) = \frac{ \sum\limits_{i=0}^{N_{\rm bl}}     w_i V(\textbf{u}_i,\nu) }{\sum\limits_{i=0}^{N_{\rm bl}} w_i},
\end{equation}
where $N_{\rm bl}$ is the number of all included baselines. A Blackman-Harris frequency taper \textbf{($H(\nu)$)} has also been applied to reduce spectral leakage in the side lobes due to the limited bandwidth.  Note that we do not normalize the beam and taper, but the steps we have taken are consistent with those taken by current 21\,cm experiments.

\subsubsection{2-D Power Spectrum }
\label{sec:calculate_ps}
\begin{figure}[ht]
\centering
\includegraphics[width=\linewidth]{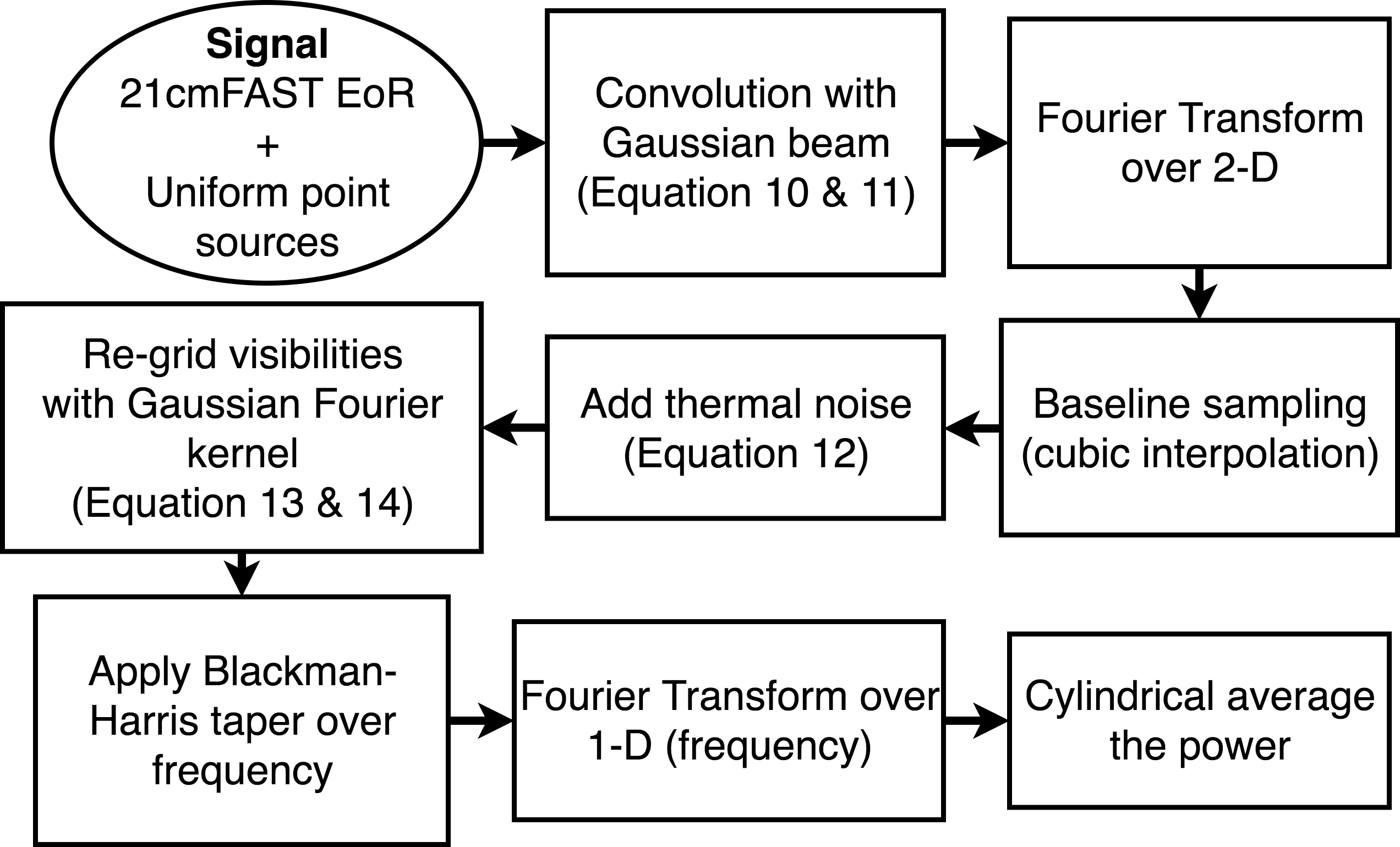}
\caption{\small\label{fig:flowchart} A summary of our instrumental algorithm. }
\end{figure}

In this work, we mostly use observational units hence $k_\perp$ and $k_\parallel$ are known as u [unitless] and $\eta$ [1 / MHz] respectively and the 2-D PS ($P(\textbf{u},\eta)$) is in unit of Jy$^2$ Hz$^2$. We compute $P(\textbf{u},\eta)$ by cylindrically averaging the power of the visibilities within radial bin \textbf{u},
\begin{equation}
\label{eqn:ps2d}
P(\textbf{u},\eta) \equiv \frac{ \sum\limits_{}^{\textbf{u}_i<\textbf{u}} V_{\rm grid}^\dagger(\textbf{u}_i,\eta)  V_{\rm grid}(\textbf{u}_i,\eta)}{\sum\limits_{}^{\textbf{u}_i<\textbf{u}} \left( {\sum\limits_{i=0}^{N_{\rm bl}} w_i} \right)}\ \ \  [ { \textrm{ Jy}^{2} \textrm{Hz}^2}],
\end{equation}
where
\begin{equation}
\label{eqn:convolved_vis}
 V_{\rm grid}(\textbf{u},\eta) =  \int V_{\rm grid}(\textbf{u}, \nu) H(\nu) \exp(-2\pi i \eta \cdot \nu) d\nu \ \ \  [ { \textrm{ Jy Hz}}].
\end{equation}The full algorithm from \S \ref{sec:instrument_component} and \ref{sec:calculate_ps} is summarized in Figure \ref{fig:flowchart}.

\subsection{Bayesian Parameter Estimation}
\label{sec:bayesian_component}
Bayesian parameter estimation through Monte Carlo Markov Chain (MCMC) is a powerful algorithm that is widely used in a variety of scientific fields to constrain parameters of interest by determining their full posterior distribution. 
We have chosen to use \cmmc, an MCMC analysis tool designed to estimate astrophysical parameter constraints from the EoR. It incorporates EoR simulation data produced on-the-fly by an optimized version of \textsc{21cmFAST} to statistically compare the models to either mock or observed data \citep{greig201521cmmc}.

The existing log-likelihood prescription used by \cmmc\ uses the 1-D PS and is given by:
 \begin{eqnarray}
\label{eqn:likelihood_old}
\ln \mathcal{L} = - \frac{1}{2} \sum_j \frac{(P_{D}(k_j) - P_M (\theta , k_j))^2}{\sigma^2_{D}(k_j)
+(\alpha P_M (\theta , k_j))^2},
\end{eqnarray}
where $P_{D}$ is the 1-D PS of the experiment or mock data, $P_M$ is the model 1-D PS, $\theta$ is the set of EoR parameters for the model, $\sigma_{D}^2$ is the uncertainty of the experiment computed using \textsc{21cmSense} \citep{pober2014next}, and $\alpha$ is a variable corresponding to the uncertainty of the model, often set to 10--20\% by inspecting how close \textsc{21cmFAST} is to radiative transfer models, with 15\% being the default value. All estimates are a function of wavenumber, $k$. In the presence of foregrounds and the instrumental beam, this prescription is only optimal\footnote{It uses all available information to generate the likelihood and therefore posterior, rather than losing some information by improper assumptions} if the following set of conditions is satisfied:
\begin{enumerate}
\item The brightness temperature fluctuations are truly Gaussian
\item The EoR signal is 3-D isotropic
\item The sources of uncertainty have 3-D isotropic signal
\item The PS modes for both the EoR and foregrounds are Gaussian
\item The PS modes are independent for both the EoR and foregrounds
\item The foreground noise is independent of the EoR parameters.
\end{enumerate}
We postulate that the first and second conditions are valid first-order assumptions and we leave the last one for future work. We discuss and address the others, along with a summary of our improvements to \cmmc, in the following sub-subsections.

\subsubsection{3-D Isotropic Uncertainties}

The 1-D PS is employed because it provides complete information about the isotropic EoR signal. However, it confounds the impact of foregrounds because they naturally reside in a larger cylindrical Fourier space (i.e., smooth-spectrum versus angular clustering). In a 1-D PS, the wedge-mode contaminates the entire annulus because the foregrounds show a non-isotropic signature, corrupting some otherwise good modes. Although some sort of avoidance method is typically used to minimize the corruption of 1-D PS modes, it does not account for the full information.

We extend the likelihood formalism by using the 2-D PS instead of the 1-D PS as it allows for the management of systematic effects arising from $k_\bot$ and $k_\parallel$ modes respectively. The 2-D PS is commonly used as an initial step to decontaminate the cosmological EoR signal from the foregrounds in the forms of the ``wedge" and the ``EoR window". 
Foreground power dominates the lower $k_\parallel$ modes, yielding higher variance in the wedge as compared to the EoR window. Using the 2-D PS means that modes in the same $|\vec{k}|$-annulus that may come from different $k_\parallel$ modes (from within and without the wedge) can be treated independently, and therefore the full information extracted.

\subsubsection{Gaussianity and Independence of Instrument-Convolved Foregrounds}
\label{sec:gaussianity}
In Eq.~\ref{eqn:likelihood_old}, the use of $\sigma_{D}^2$ follows the assumption that the input data are Gaussian and independent. While this will hold true for the EoR signal since the cosmological signal is close to Gaussian, in the presence of foregrounds and instrumental effects, this assumption might break. We will focus on investigating the Gaussianity and independence of the instrument-convolved foregrounds because the variance of the data is dominated by the variance of the foregrounds in most bins. This can then dictate the appropriate formalism to adopt, without adding unnecessary complexity. Since we have established in the previous sub-section that we will be using the 2-D PS to accommodate the presence of foregrounds, we will continue this discussion using the 2-D PS of instrument-convolved foregrounds that were simulated based on our models; the ensemble is then used to compute the skewness and correlation in each 2-D mode. 

\begin{figure}[ht]
\centering
\includegraphics[width=\linewidth]{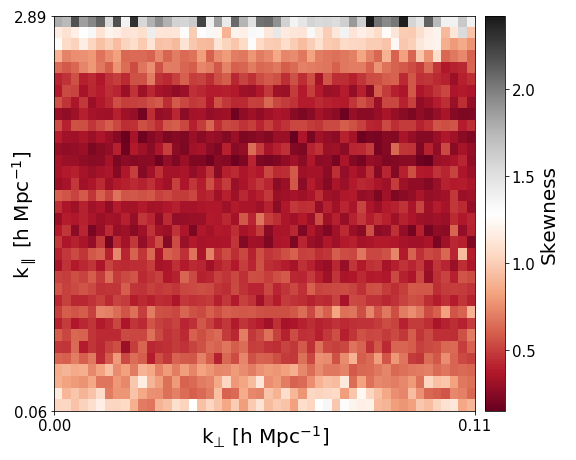}
\caption{\small\label{fig:gaussian_PS} The skewness of the 2-D PS of the instrument-convolved point source foregrounds. The bins are predominantly close to Gaussian over the primary range of interest (0.1 to 1 $h$ Mpc$^{-1}$). }
\end{figure}

We investigate the Gaussianity of the 2-D PS by finding the skewness in bins of $k_\bot$ and $k_\parallel$. The skewness measures the asymmetry of the probability distribution about its mean, with a skewness of 0 being perfectly symmetric and increasing skewness showing increasing asymmetry. Figure \ref{fig:gaussian_PS} shows the skewness of the 2-D PS of the instrument-convolved point source foregrounds. Over the primary range of interest (0.1 to 1 $h$ Mpc$^{-1}$), the bins are predominantly close to Gaussian. Regions of higher skewness are found only at either high $k_{||}$, where the signal is low, or low $k_{||}$, where the foregrounds are strong so the mode is therefore heavily down-weighted in the likelihood, and thus in our fits for the underlying parameters $\theta$. We are thus content to maintain the assumption of a Gaussian likelihood.

\begin{figure*}[ht]
\centering
\includegraphics[width= \textwidth]{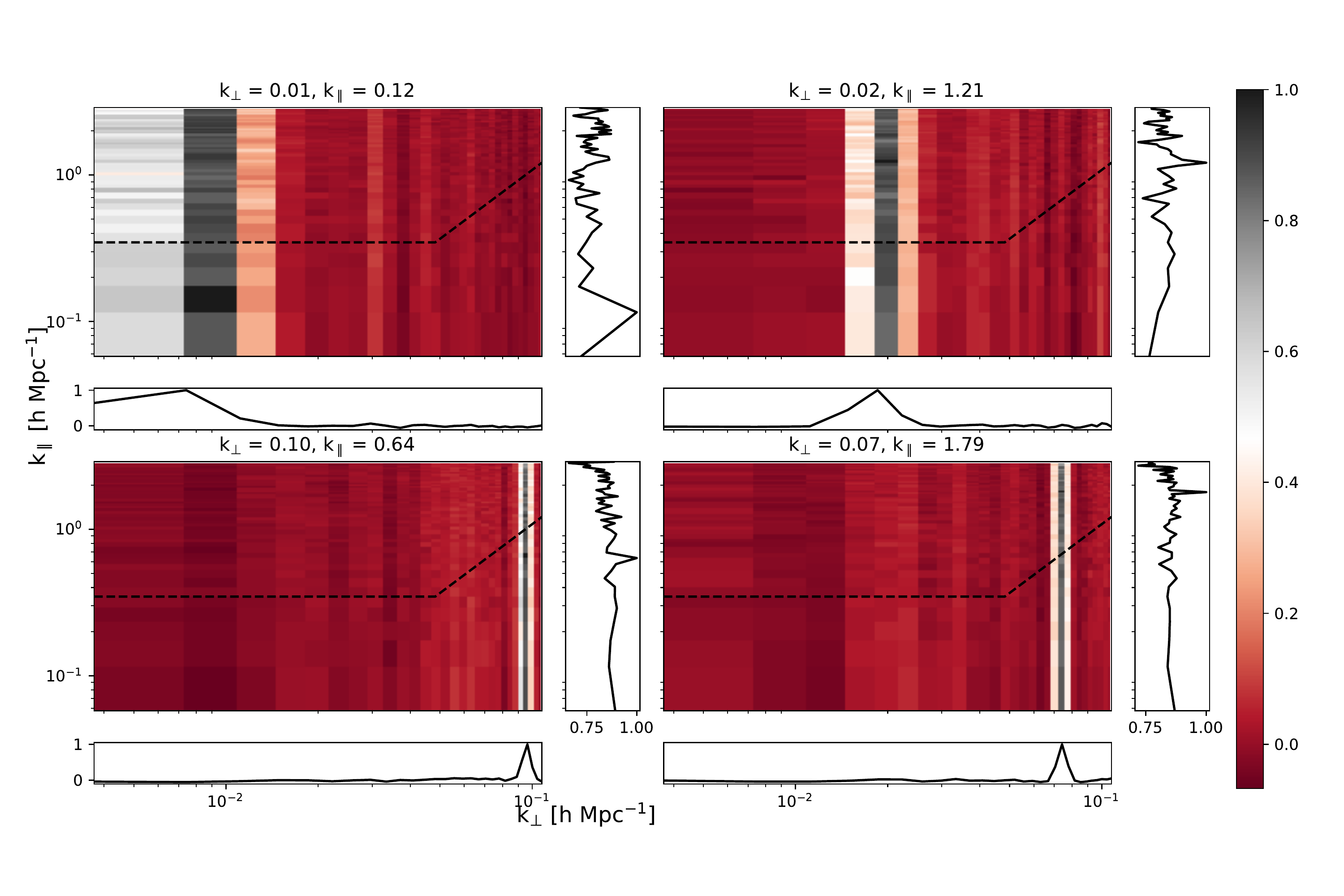}
\caption{\small\label{fig:corr_PS} The correlation coefficient of the 2-D PS of the instrument-convolved foregrounds between a single cell with all other cells. The horizontal and vertical plots show cuts through the 2-D space at the chosen single cell locations, demonstrating the degree of correlation in the angular and line-of-sight modes. Each chosen ($k_\bot, k_\parallel$) bin is completely correlated with itself (black cell in main panels corresponding to highest point in the smaller side panels). However, also note that the bin is highly correlated over $k_\parallel$ and the correlation spills over to the two $k_\bot$ bins adjacent to the chosen bin.}
\end{figure*}

We investigate the assumption of independence of the instrument-convolved foreground model by finding the correlation coefficient between $k_\bot$- and $k_\parallel$-bins in the 2-D PS. The correlation measures the strength of the joint variability of two random variables, in this case the bins of $k_\bot$ and $k_\parallel$, with possible values ranging from -1 being completely anti-correlated to +1 being completely correlated. As with Gaussianity (cf. \S\ref{sec:gaussianity}), we focus on the foregrounds rather than the 21\,cm signal.

Figure \ref{fig:corr_PS} shows the correlation coefficient of the 2\nobreakdash-D PS of the instrument-convolved foregrounds of four individual ($k_\bot, k_\parallel$) bins with all other bins. 
The right- and bottom- sub-panels of each major panel show the correlation coefficient along a 1-D slice (in the displayed dimension) through the selected 2-D bin of each main panel. The dotted black line divides the area between the wedge and the EoR window. As expected, each chosen ($k_\bot, k_\parallel$) bin is completely correlated with itself (black cell in main panels corresponding to highest point in the smaller side panels). However, we can also see that the bin is highly correlated over $k_\parallel$ and the correlation spills over to the two $k_\bot$ bins adjacent to the chosen bin. This is due to the frequency-dependence of the instrument, resulting in chromatic mode-mixing of foregrounds \citep{morales12,datta10,vedantham12,trott2012impact}. As such, we can conclude that the 2-D PS of the instrument-convolved foregrounds is not independent between modes and that we must use the covariance, instead of the variance, in the likelihood prescription. Neglecting the dependence of the modes may result in over-constraining of parameters, regardless of whether the PS is is 1-D or 2-D. \\

We thus extend the likelihood prescription to use the multivariate normal likelihood, $\ln \mathcal{L}, $ following the use of the 2-D PS and the presence of foregrounds. $\ln \mathcal{L} $ is given by
\begin{align} 
\label{eqn:likelihood_new}
\ln \mathcal{L} =  -\frac{1}{2} \left[\left(P_D - P_M( \vec{\theta})\right)^T 
\Sigma^{-1} \left(P_D - P_M( \vec{\theta})\right)\right],
\end{align}

where $\Sigma$ is the covariance\footnote{$\Sigma$ is computed numerically using the Monte Carlo method to simulate the noise and foreground over 1000 random realizations at the fiducial parameters. The ensemble of realizations is passed through the algorithm summarized in Figure \ref{fig:flowchart} to produce the 2-D PS and calculate $\Sigma$. This is pre-computed prior to the MCMC. We note that the covariance may not have converged (see e.g \cite{dodelson2013effect} and \cite{taylor2014estimating}) hence further investigation is necessary, but we have left this for future work.} of our \textit{fiducial} noise and foreground ($N+FG$) model with the addition of sample variance (second term of following equation), so that the total variance, $\sigma^2_{T}$, is given by
\begin{equation}\label{eqn:variance}
    \sigma^2_T = \sigma^2(P_{N+FG;M}) + \frac{P_{21;M}^2}{ \sum_j (\sum\limits_{i=0}^{N_{\rm bl}} w_i)^2};
\end{equation}
\begin{equation}
    P_D = P_D(k_\bot, k_\parallel) = P_{21+N+FG;D}
\end{equation}
is the (mock) data PS, assumed to be comprised of a 21\,cm component, and a noise and foreground component;
and
\begin{equation}
    P_M(k_\bot, k_\parallel|\vec{\theta}) = P_{21} (k_\bot, k_\parallel|\vec{\theta}_{21}) + P_{N+FG} (k_\bot, k_\parallel| \vec{\theta}_{N+FG})
\end{equation}
is a parameter-dependent \textit{model} of the data. Note that we use the \textit{expected} power of noise and foregrounds instead of just one specific realization for the model. Because it is computationally-expensive and difficult to calculate the expected power of all components at each iteration, we have modelled the 21\,cm signal separately and use the same expected foregrounds and noise power each time, which is possible because we do not have $\theta_{N+FG}$ parameter in the likelihood. This also assumes that the cross-power\footnote{An explicit derivation of its variance is shown later in \S \ref{sec:results} (Equation \ref{eqn:var-residual})} terms are negligible and is a reasonable first-order approximation consistent with other works; however, we will see in \S \ref{sec:results} that this assumption is not true.

Note that $P_D$ and $P_M$ are \textit{not} the true power spectrum computed from the simulation box. They are related to the true power spectrum, $P^*$, by $P=\hat{W} P^*$ where $\hat{W}$ is all the window functions that encompasses the instrumental and analysis effects such as the beam and frequency taper used in this research. The fact that we compute $P_D$, $P_M$, and $\Sigma$ using exactly the same pipeline means that they are self-consistent hence the effects of $\hat{W}$ should cancel out in the likelihood. Also note that the form of Eq.~\ref{eqn:likelihood_new} explicitly assumes that the covariance does not depend strongly on the astrophysical parameters. Otherwise, it would require an extra term of $-\ln {\rm det}(\Sigma^{-1})$. This is a reasonable assumption for most cylindrical $k$-modes which are dominated by foregrounds whose parameters we are not directly interested in.

\subsection{Extensions to \cmmc}

The foregrounds and instrumental components, along with other improvements developed in this work and described in the previous sections, are combined in a publicly-available plug-in to the new \cmmc, called \textsc{py21cmmc-fg}\footnote{Found at \url{https://github.com/BellaNasirudin/py21cmmc-fg}}. It makes use of \cmmc's new ability to allow the user to arbitrarily insert code to modify the lightcone before producing a likelihood. This creates realizations of the EoR signal obscured by the foregrounds which are measured by an instrument model and then used as input to the log-likelihood.

To summarise, the extensions to \cmmc\ that are available in \textsc{py21cmmc-fg} include:
\begin{itemize}
\item \textbf{a foreground model:} includes both diffuse and point-source foregrounds (although only the latter is used in this work for simplicity)
\item \textbf{an instrument model:} includes a Gaussian beam model, the array baseline sampling and thermal noise
\item \textbf{calculation of the 2-D PS:} to separate the cosmological EoR signal from the foregrounds
\item \textbf{the covariance of the PS:} to account for the correlated instrument-convolved foregrounds
\item \textbf{the likelihood prescriptions for the MCMC:} expand to multivariate normal distribution
\item \textbf{stitching of the lightcones:} to account for the wide FOV of the instrumental beam model
\item \textbf{padding of the sky:} to increase the image resolution for Fourier Transform
\item \textbf{Fourier beam gridding kernel:} to properly interpolate, average and weight visibilities onto a grid
\end{itemize}
All of these have been discussed in detail in this section, except for the stitching of the lightcones. We will discuss this in the next section (\S \ref{sec:EoR_analysis}) when we explore the observational effects on the EoR lightcones. 

\subsection{Comparison to Existing Framework}

The final focus of this section is to compare the uncertainty level from our framework,  \textsc{py21cmmc-fg}, to that from the publicly-available existing framework, \textsc{21cmSense} \citep{pober2013baryon,pober14}. In general, \textsc{21cmSense} performs a separate calculation of the instrument sensitivity arising from thermal noise and cosmic variance while avoiding the foreground-dominated region; the uncertainty file is then used as input into \cmmc\,. For more details, we refer readers to  \cite{pober2013baryon} and \cite{pober14}.

For \textsc{21cmSense}, we adopt the drift scan mode and ``moderate" foreground removal with the MWA Phase II baselines, $A_{\rm eff}= 21$ m$^2$ and $T_{sys}= 240$ K. We also use $\Delta t = 2 \times 10^5$ hours in both frameworks. The reasoning behind this value is explained in detail in \S \ref{sec:mcmc_analysis}, but in general, we use this value so that we can approximate the noise level for 1000 hours observation of the future SKA\_LOW1.

Note that the uncertainty with sample variance is calculated differently in the two frameworks. Our framework calculates $\sigma^2_T$ based on Equation \ref{eqn:variance}, while \textsc{21cmSense} defines $\sigma^2_T$ as
\begin{align}
\label{eqn:variance_cmsense}
    \sigma^2_{T} (k) = \left( \sum_i  \frac{1}{(P_{21}(k) + P_{N, i}(k))^2}\right)^{-\frac{1}{2}},
\end{align}
whereby an inverse-weighted summation is performed over the $uv$ cell $i$ for each $k$ mode bin \citep{pober2013baryon}. The main difference between the two frameworks is the presence of the cross-terms between the EoR PS and the noise variance for \textsc{21cmSense}.

\begin{figure}[]
\centering
\includegraphics[width= 0.5\textwidth]{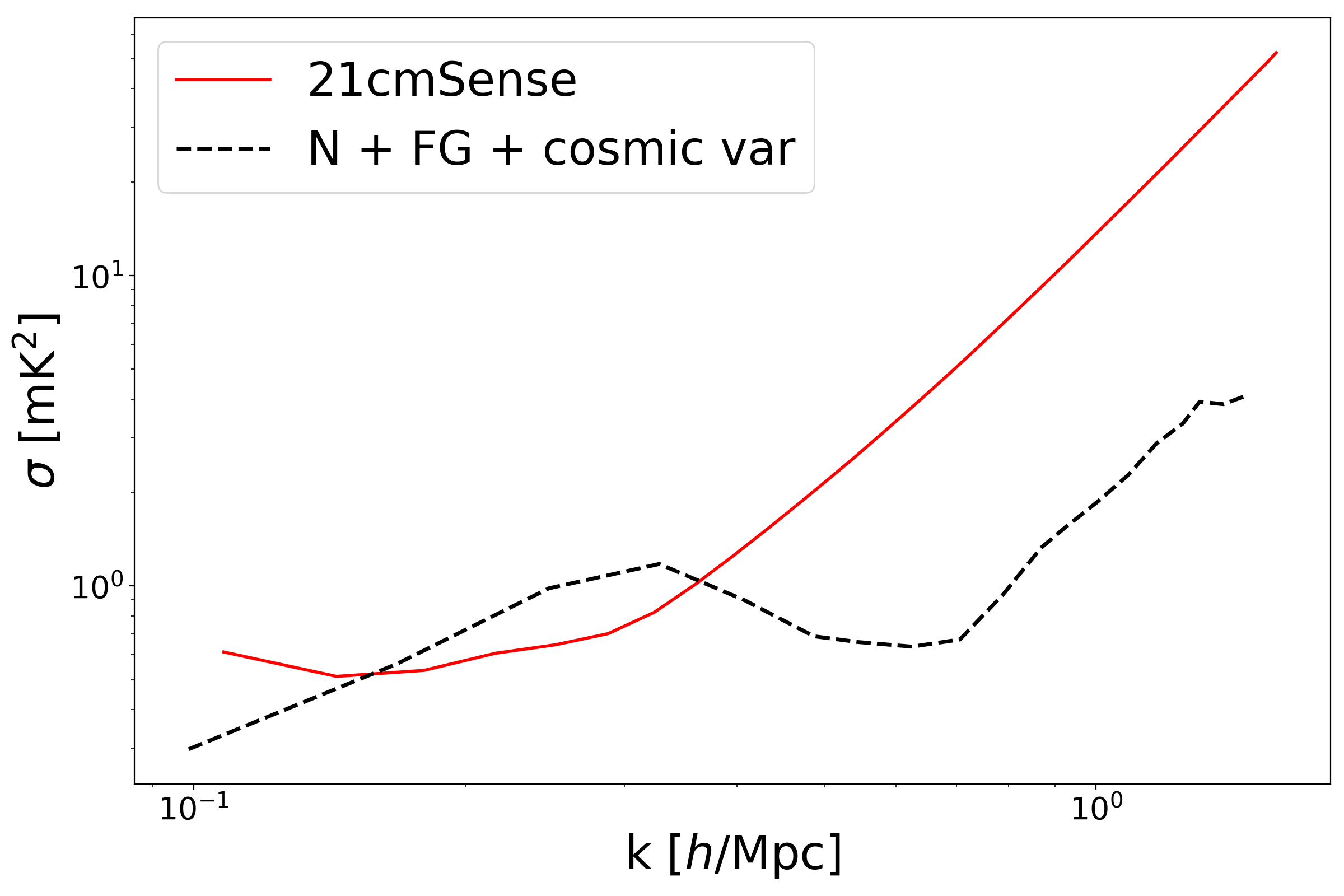}
\caption{\small\label{fig:noise-std_comparison2} The total uncertainty from \textsc{21cmSense} (red solid line) compared to the total uncertainty from our framework (black dash line) with $T_{sys}=240 $ K and $\Delta t =$ 2$\times 10^5$ hours.}
\end{figure}

In addition to the calculation of sample variance, \textsc{21cmSense} and \textsc{py21cmmc-fg} have a few key differences that can affect the total uncertainty level. We have summarized these differences in Table \ref{tbl:differences}. The resulting total uncertainty from the two frameworks are presented in Figure \ref{fig:noise-std_comparison2}. The total uncertainties loosely match at the relevant modes where the signal dominates, with a difference of less than half an order of magnitude. At high $k$, however, the difference is as high as one order of magnitude.  We attribute the difference in total uncertainty to the missing Earth rotation synthesis which essentially fills up more $uv$ space hence increasing the overall noise level (i.e., less modes can combine coherently). Even though the Fourier beam gridding kernel we employ partly takes this into account, it does not restore the noise to the full level that it would be with a proper rotation synthesis. We note that this is a shortcoming of our framework, and is left for future work.

\begin{table}[] 
\begin{tabular}{l|l|l|}
\cline{2-3}
                                                     & \textbf{21cmSense} & \textbf{py21cmmc-fg} \\ \hline
\multicolumn{1}{|l|}{PS}                             & 1-D                & 2-D                  \\ \hline
\multicolumn{1}{|l|}{Frequency-dependent baselines}  & No                 & Yes                  \\ \hline
\multicolumn{1}{|l|}{Cross-power in variance} & Yes                & No                   \\ \hline
\multicolumn{1}{|l|}{Foregrounds}                    & Avoidance          & Suppression            \\ \hline
\multicolumn{1}{|l|}{Earth rotation synthesis}       & Yes                & No                   \\ \hline
\multicolumn{1}{|l|}{Baseline sampling on EoR PS}    & No                 & Yes                  \\ \hline
\multicolumn{1}{|l|}{Frequency taper}                & No               & Yes                  \\ \hline
\multicolumn{1}{|l|}{Gridding kernel}                & No              & Yes                  \\ \hline
\end{tabular}
\caption{\small \label{tbl:differences} The differences between \textsc{21cmSense} and \textsc{py21cmmc-fg}.}
\end{table}

\section{Observational Effects on EoR Lightcones}\label{sec:EoR_analysis}

\begin{figure}[ht]
\centering
\includegraphics[width= 0.5\textwidth]{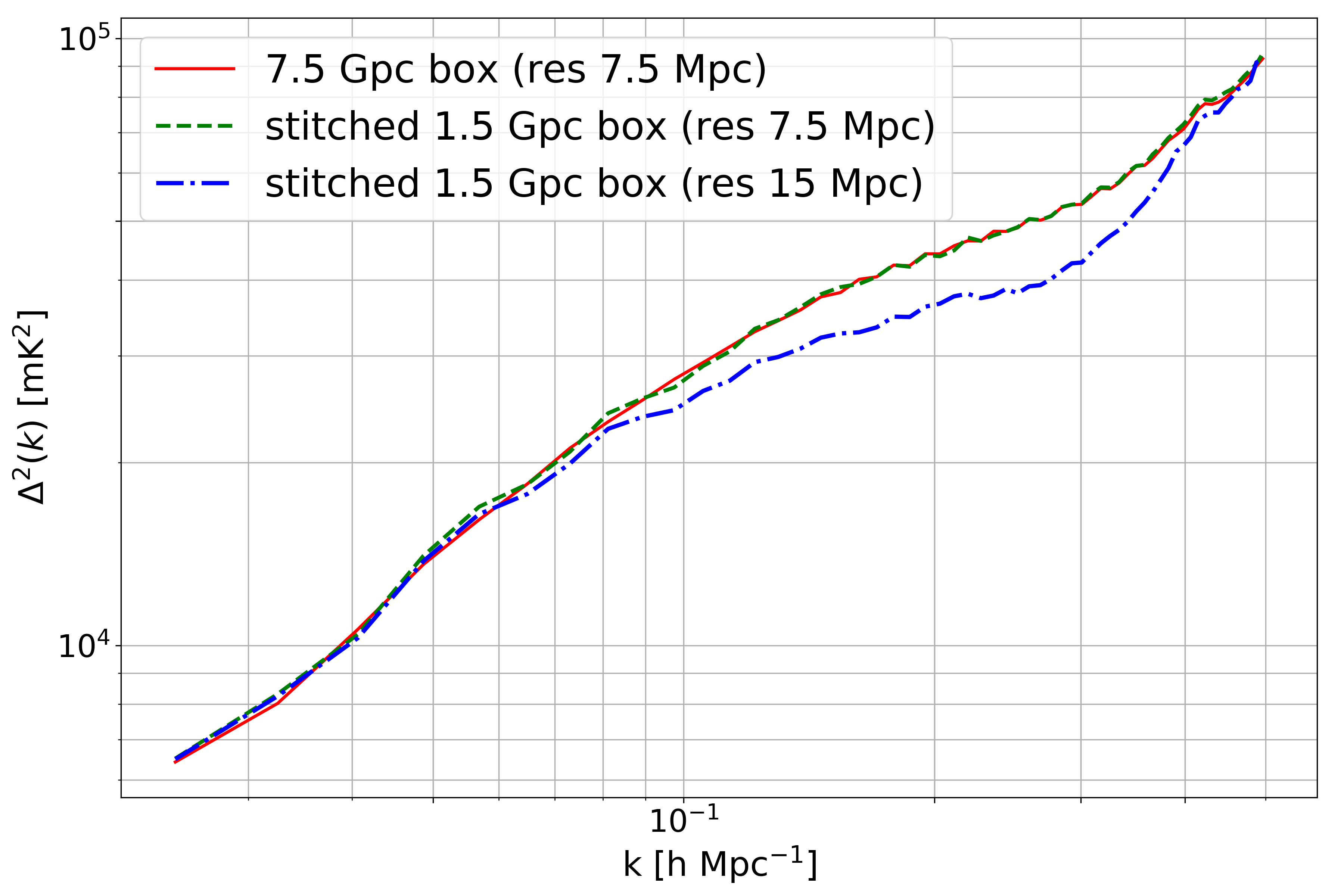}
\caption{\small\label{fig:1D_compare} Comparison of the 1-D PS of the 7.5 Gpc box (red), the stitched 1.5 Gpc box (green dotted), and the stitched 1.5 Gpc coarsened to a resolution of 15 Mpc (blue dash-dot). The 7.5\,Gpc box and the stitched 1.5\,Gpc box agree very well on most scales, but the 1-D PS of the stitched and coarsened box differs from the 1.5\,Gpc box at high $k$ due to finite resolution.}
\end{figure}

In this section, we explore the observational effects of adding foregrounds to the EoR signal. This is first done by explaining the steps we have taken to ensure that both the EoR signal and foregrounds can be added together. We then analyse the effects on the 1-D and 2-D power spectra and compare the differences in image space.

In order to add the foreground and EoR lightcones, they need to cover the same angular size and have the same angular resolution. At $z=6$, the resulting angular sky size of a 1 Gpc$/h$ simulation box, which is considered to be large in theoretical studies, is merely $\sim 0.12$ radian. On the other hand, the FOV of the MWA is $\sim 1$ radian. Running a massive box that covers the MWA FOV is unrealistic and computationally expensive, so instead, we have made a realization of the lightcone across the mock sky by assuming that the same structure is periodically repeated. To preserve the wavenumber, we have ensured that the full box is used in the stitching i.e., we use an integer number of boxes. We also limit ourselves to using scales smaller than the box size in the parameter estimation to avoid systematics due to this repetition.

Furthermore, due to memory-limitation, we have opted to coarsen the EoR signal using interpolation. Of course, the coarsening process can be avoided by setting the EoR lightcone from \textsc{21cmFAST} to be of the same angular size. This, however, will affect the small-scale processes involved in determining $\delta T_b$, so we prefer to perform the coarsening post-stitching.

To study the effects of tiling and coarsening of the lightcone, we use simulation boxes of size 1.5 and 7.5\,Gpc, both with a resolution of 7.5\,Mpc. 
While we will typically work with the 2-D PS, we use the 1-D PS when comparing data quantitatively in this section. 
We compare the 1-D PS of the 7.5\,Gpc box, the stitched\,1.5 Gpc box, and the stitched\,1.5 Gpc coarsened to a resolution of 15 \,Mpc in Figure \ref{fig:1D_compare}.  The $y$-axis shows the dimensionless 1-D PS while the $x$-axis shows the $k$ scales.

The 7.5\,Gpc box and the stitched 1.5\,Gpc box agree very well on most scales, suggesting that stitching the box does not negatively affect the PS. However, the 1-D PS of the stitched and coarsened box differs from the 1.5\,Gpc box at high $k$\footnote{The curves converge again at high $k$ because modes $\geq 0.2$ $h/$Mpc (corresponding to the Nyquist value) is not reliable.}. This is, of course, expected as the coarsening of the box results in a different resolution, hence affecting the small scale structure of the $\delta T_b$. The stitching of boxes and coarsening of the resolution may potentially affect the statistics of the PS due to lack of cosmic variance. 

Herein, we have been presenting plots in cosmological units. In the code and hereafter in the paper, we have opted for observation units of Jy, Hz and radians to minimise inconsistencies and errors in converting between the two units. Note that the choice of units is inconsequential for the likelihood, as long as the same units are used in numerator and denominator.

\section{MCMC Analysis}
\label{sec:mcmc_analysis}

\begin{figure*}[]
\centering
\includegraphics[height= 0.42\textheight]{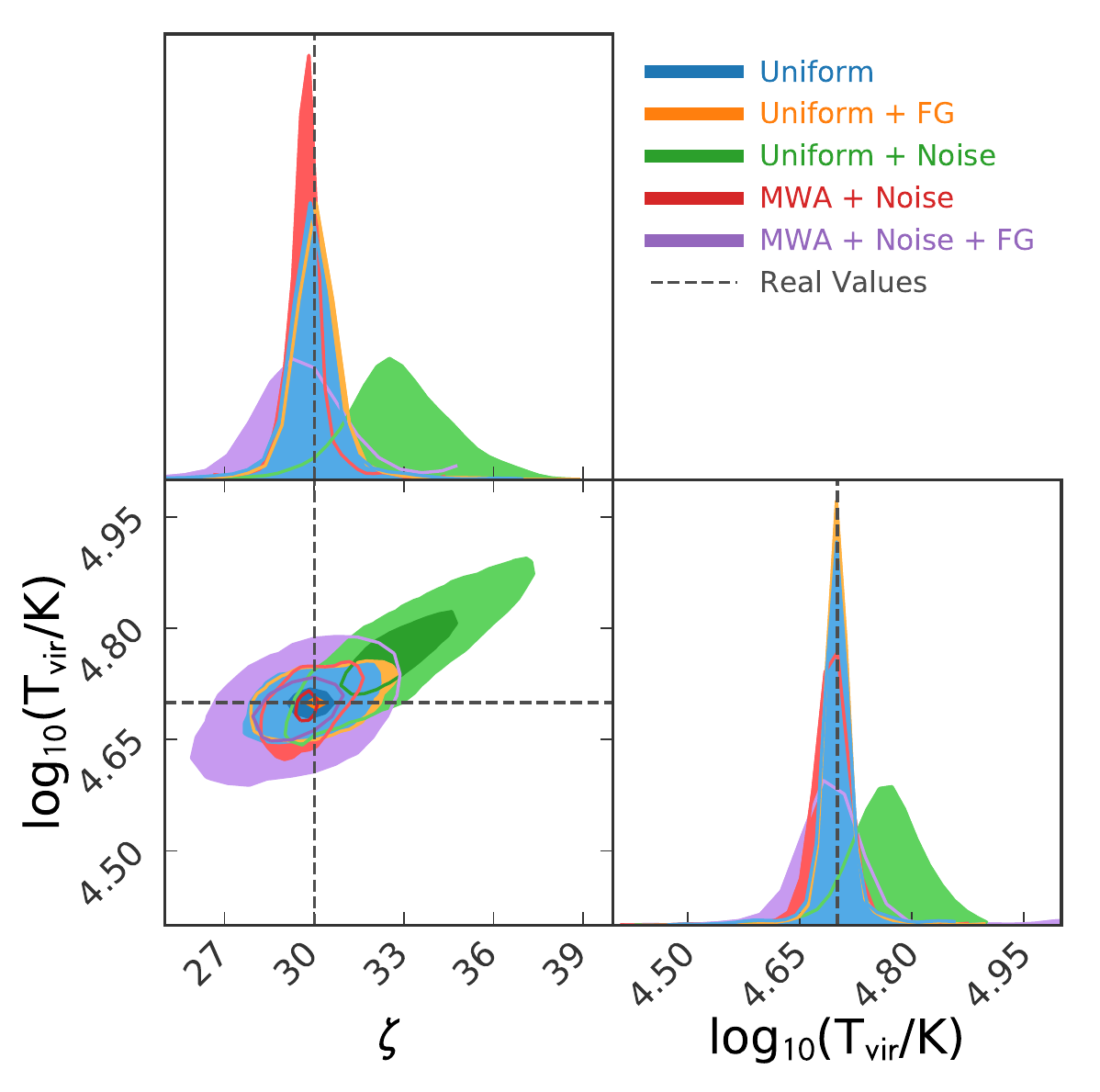}
\caption{\label{fig:corner_steps} A corner plot comparing the constraints our systematic tests: uniform $uv$ sampling with no thermal noise or foregrounds (blue); uniform $uv$ sampling with point-source foregrounds but no thermal noise (orange); uniform $uv$ sampling with 1000 hours of thermal noise with the MWA but no foregrounds (green); MWA baseline sampling with $2 \times 10^5$ hours of thermal noise with the MWA but no foregrounds (red); MWA baseline sampling with $2 \times 10^5$ hours of thermal noise with the MWA and point-source foregrounds (purple). The use of $2 \times 10^5$ hours of thermal noise with the MWA is comparable to the noise level of 1000 hours with the SKA. The artificial ``uniform" $uv$ sampling unnecessarily inflates the impact of thermal noise (green) as it restricts the number of baselines per $uv$ cell to one and the small collecting area of the MWA gives a signal-to-noise ratio of order unity; both of these factors results in a biased estimate. With the addition of baselines sampling and the use of SKA sensitivity, the constraints are remain unbiased (purple and red) although the final test (red) gives a really inflated constraint compared to the other tests. Note that these constraints still have the burn-in phase included. }
\end{figure*}

For parameter estimation, we use an EoR lightcone of size 750\,$h^{-1}$Mpc with 250 cells. The lightcone is stitched together to cover 1 steradian of the sky and coarsened to 800 cells (resolution of $\approx$ 24\,$h^{-1}$Mpc) before being padded by zeros to 3 times the sky size to increase the resolution for the Fourier Transform. The bandwidth of the observation is 10\,MHz from 150 to 160\,MHz with 100 frequency channels. We assume that $T_{sys}=240$\,K, corresponding to the average sky temperature at 150\,MHz.

To ensure that our MCMC framework is consistent, we first perform simpler tests in which we attempt to loosely constrain only two parameters, $\zeta$ and $T_{vir}$ (with fiducial values of $\zeta$=30.0 and $T_{vir}=4.7$ \,log$_{10}$K). These systematic tests include:
\begin{enumerate}
\item Uniform -- regular and filled -- $uv$ sampling with no thermal noise or foregrounds
\item Uniform $uv$ sampling with point-source foregrounds but no thermal noise
\item Uniform $uv$ sampling with thermal noise but no foregrounds
\item MWA baseline sampling with thermal noise but no foregrounds
\item MWA baseline sampling with thermal noise and point-source foregrounds
\end{enumerate}
Step 5 constitutes the end-goal of this work (but with all four parameters from Table~\ref{tbl:params}). Note that with these tests, we are not interested in properly estimating the values of the parameters; we only want to gauge whether the MCMC walkers would assemble around the right values, thereby the burn-in phase of the walkers are still included in this section. We initially set the observation time to 1000 hours consistent with other research, but the observation time is later increased to 2$\times 10^5$ hours (the motivation of which is explained later in this section). To clarify, only steps 1 - 3 are done with 1000 hours of observation time. The final two steps, along with everything else after, are done with 2$\times 10^5$ hours.

We present a corner plot of our checks in Figure \ref{fig:corner_steps}, which shows the projection of the values of the parameters over 1-D and 2-D space density. The black dash lines show the actual value of the parameter while the colored, shaded regions show the values of the estimated parameter that have been projected over the density space in 1-D (diagonal plots) and 2-D (non-diagonal plots).

It is clear that convolving the stitched and coarsened EoR signal with the Gaussian beam and implementing the 2-D PS and the multivariate normal distribution in the likelihood prescription do not negatively impact the parameter estimation, as apparent from the constraints from Step 1 (blue region). The addition of point source foregrounds in Step 2 (orange) does not significantly expand the constraints. This can be attributed to the fact that the foreground contamination is fully contained at smaller $\eta$ due to the uniform $uv$ sampling.

\begin{figure*}[]
\centering
\includegraphics[height= 0.5\textheight]{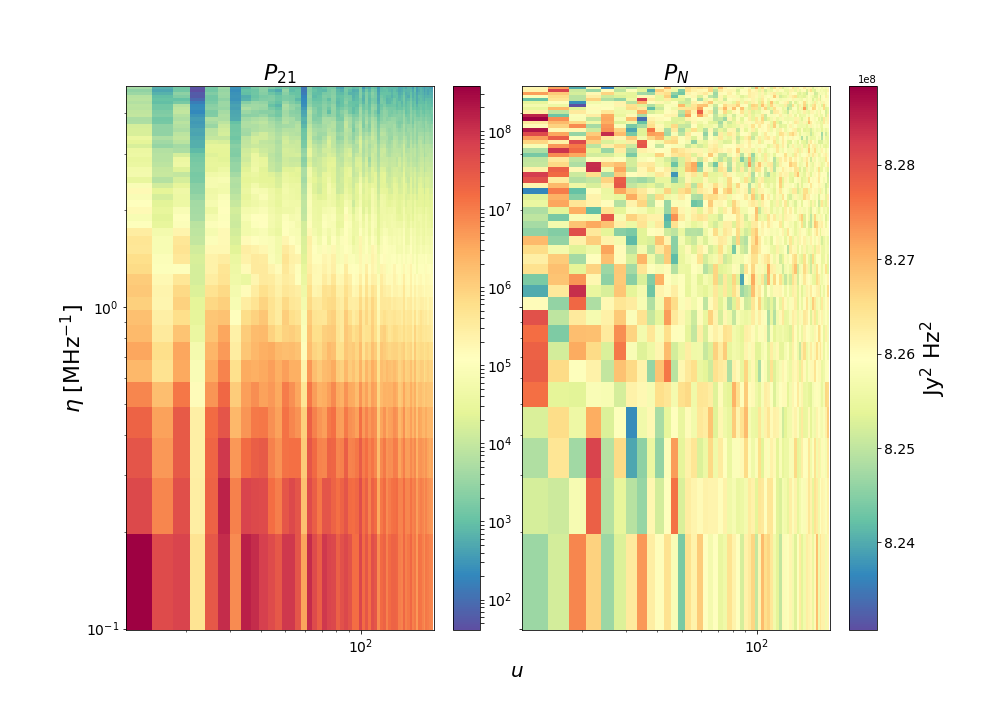}
\caption{\small\label{fig:ps_eor_noise} The power spectra of the EoR signal (left) and the thermal noise with the MWA sensitivity (right) in unit of Jy$^2$Hz$^2$. With the uniform $uv$ sampling, the signal to noise ratio is of order unity for 1000 hours of observation time due to the small collecting area of the MWA. While this is enough for an EoR signal detection over the slope of the 1-D PS, it is not sensitive enough for EoR parameter estimation with the MWA. }
\end{figure*}

However, with the addition of thermal noise in Step 3 (green region), the peak of the distribution is significantly shifted with respect to the true value, indicating that the results are biased. We note also that the artificial ``uniform" $uv$ sampling unnecessarily inflates the impact of thermal noise, as it restricts the number of baselines per $uv$ cell to one. To investigate the cause of the bias, we plot the 2-D PS of the EoR signal and the thermal noise, presented in Figure \ref{fig:ps_eor_noise}. With the uniform $uv$ sampling, the signal-to-noise ratio is of order unity for 1000 hours of observation time due to the small collecting area of the MWA, either with the MWA baselines sampling or the uniform $uv$ sampling. While this is enough for an EoR signal detection over the slope of the 1-D PS, it is not sensitive enough for EoR parameter estimation with the MWA.

To enable parameter constraints for the remainder of the work, we henceforth increase $\Delta t$ in Equation \ref{eqn:thermal_noise} to 2$\times 10^5$ hours for the MWA-based instrument model that we have. This $\Delta t$ is equivalent to adopting an SKA-like sensitivity for which $A_{\rm eff}= 300\, {\rm m}^2$ but is achievable with only 1000 hours.
As seen in Step 4 (red region), both the use of the larger collecting area and the presence of $uv$ sampling reduces the noise, resulting in the MCMC converging to the model values.

In Step 5 (purple region), the presence of point source foregrounds expands the posteriors by 100\% compared to thermal noise alone without bias. The reason the purple contour is larger than the red, while the orange is not larger than the blue, is that non-uniform sampling implies chromaticity which creates the wedge, and therefore removes a lot of otherwise useful modes.

\section{Parameter Estimation Results}
\label{sec:results}

\begin{figure}[]
\centering
\includegraphics[width= 0.5\textwidth]{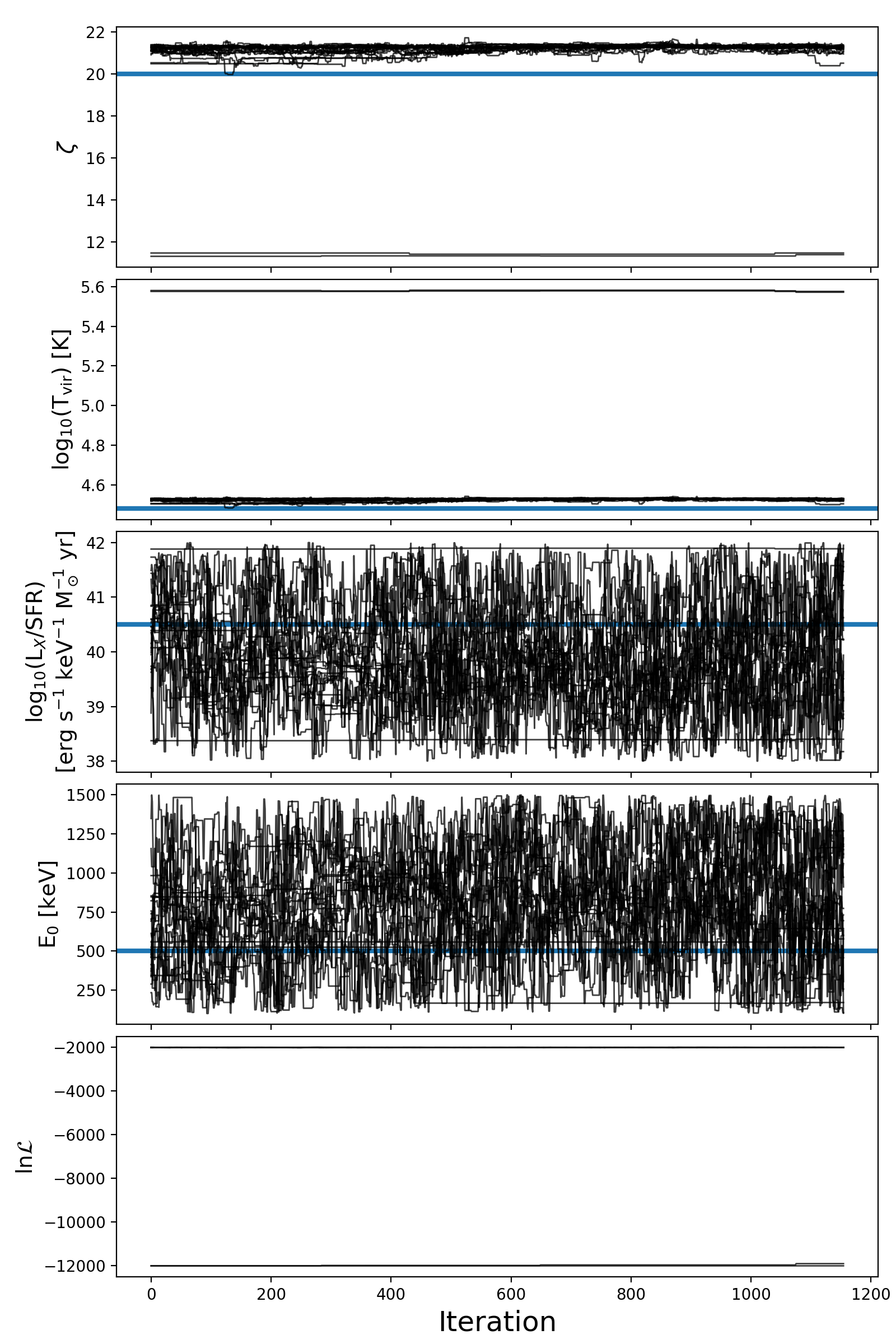}
\caption{\small\label{fig:trace} An example trace plot from the 160--170 MHz band with thermal noise and foregrounds. The blue lines show the true value of the parameter while the black lines show the progression of the parameter estimation at each iteration. In this particular case, we have ignored the walkers with $\ln \mathcal{L}<-4000$.}
\end{figure}

Now that we have established the robustness of our additional model complexities, we can use the MWA baseline sampling with thermal noise and point-source foregrounds to constrain the four reionization parameters in Table \ref{tbl:params}. We use three different frequency bands: 150--160, 160--170, and 170--180\,MHz to observe the EoR signal, which is obscured by point-source foregrounds in the presence of instrumental effects and thermal noise.

Before we proceed with our results, we would like to emphasize that our pipeline is computationally intensive and time-consuming. With 5 chains run for each parameter, we have 20 walkers in total. One iteration per walker uses 10 GB of memory per node and takes roughly 20 minutes to completely simulate the EoR brightness temperature field, tile and coarsen it to cover the whole sky, add a Gaussian beam, Fourier Transform the beam-convolved sky and sample the $uv$ sample, re-grid the visibilities and cylindrically average them to find the 2-D PS before finding the log-likelihood. The bottleneck of our pipeline is in re-gridding the visibilities using a Gaussian beam kernel after baseline sampling, which takes around 15 minutes to complete.

To test the performance, we use a Gaussian Process model to fit for the maximum likelihood autocorrelation function which is used to estimate the autocorrelation time, $\tau$. We found that $\tau \approx 200$ iterations irrespective of which frequency band is being used, with the recommended total iteration being 50$\tau$. However, because of the memory and time limitations we have mentioned above, we have only run our chains up to around $5\tau$. This is not enough to definitively claim that our chains have properly converged, but the results we currently have are useful to give an initial estimate of how well we can constrain the EoR parameters in the presence of foregrounds and thermal noise with our pipeline and gauge the plausibility of our assumptions.

Furthermore, in calculating our constraints, we have ignored walkers that are stuck in local minima. An example of this is shown in Figure \ref{fig:trace}, where we have plotted the trace plots which show the values of the four EoR parameters at every iteration and the corresponding $\ln \mathcal{L}$ after burn-in for the 160--170 MHz band with thermal noise and foregrounds. In this particular case, we have ignored the walkers with $\ln \mathcal{L}<-4000$.

\begin{figure*}[]
\centering
\includegraphics[height= 0.55\textheight]{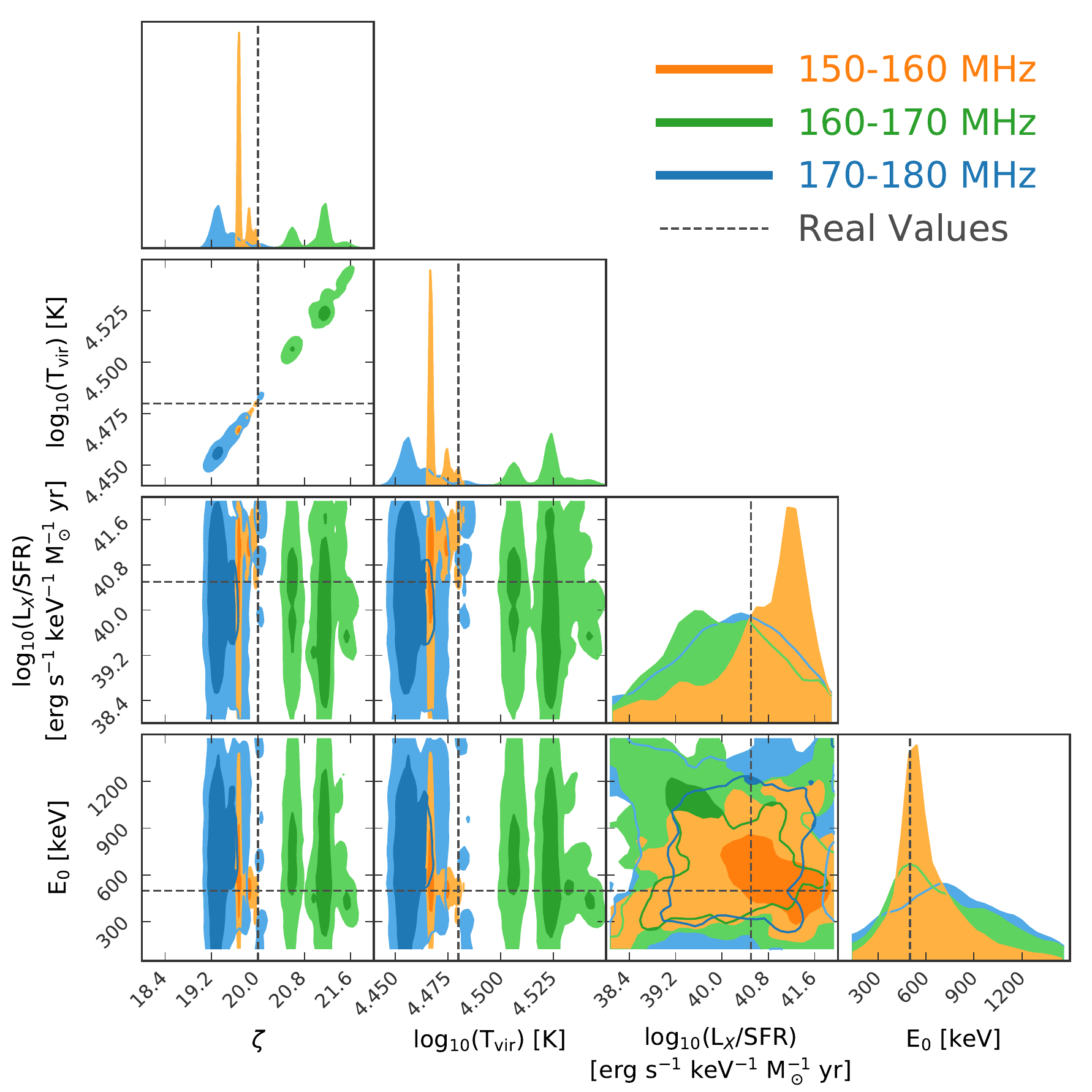}
\caption{\label{fig:final_corner-3obs-noisefg} A corner plot comparing the constraints of EoR parameters in the presence of instrumental noise and foregrounds in the different frequency bands: 150--160 (orange), 160--170 (green), and 170--180 (blue) \,MHz. }
\end{figure*}

\begin{figure*}[]
\centering
\includegraphics[height= 0.3\textheight]{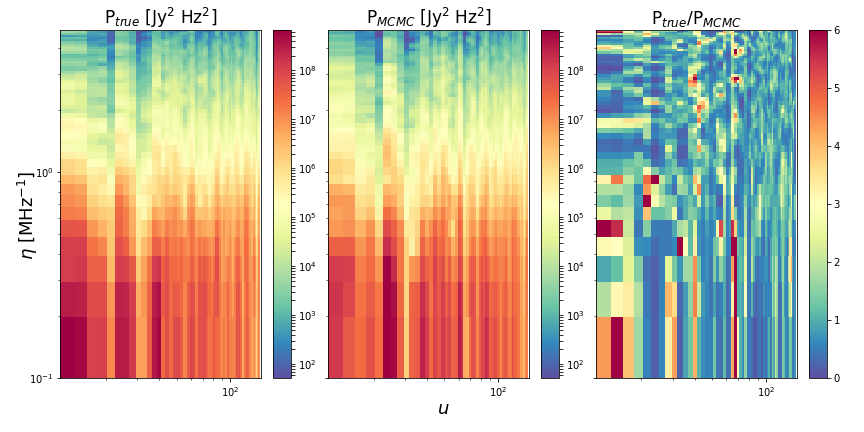}
\caption{\small\label{fig:ps_eor_params} The 2-D power spectra of the EoR signal (i.e. without foregrounds or thermal noise) in the 160--170 MHz band for the true values (left) and the MCMC values (middle) -- the median values of $\zeta=21.13$, log$_{10}(T_{vir})=4.52$ , L$_X$/SFR = 40.00, and E$_0$=0.656 keV-- along with their ratio. The true parameters give a maximum power that is up to 6 times higher than the MCMC parameters in the EoR window, suggesting that it is deceptively easy to discriminate between the two models. }
\end{figure*}

We present a corner plot of the different observation bands in Figure \ref{fig:final_corner-3obs-noisefg} in which the orange, green, and blue shaded regions represent the 150--160, 160--170, and 170--180\,MHz bands respectively. The black dashed lines show the true values of the parameters ($\zeta=20$, log$_{10}(T_{\rm vir})=4.48$ , L$_X$/SFR = 40.5, and E$_0$=0.5 keV). The darker/lighter shaded regions represent 1$\sigma$/2$\sigma$ confidence regions corresponding to the 84$^{th}$ and 16$^{th}$ percentiles. From the figure, it is clear that L$_X$/SFR and $E_0$ are relatively unconstrained in all frequency bands and that $T_{\rm vir}$ is tightly constrained. Note that  $\zeta$ and $T_{vir}$ are quite correlated and biased in all bands. This is especially true for the 160--170 MHz band in which the green region is further from the truth values. The bias suggests that the primary effects of $\zeta$ and $T_{vir}$ are on regions which are most affected by foregrounds on the larger scales. A bias indicates either a very anomalous noise realization (more than 2 sigma over the whole 2D parameter space), or an incorrect model.

To investigate the reason for the bias in the presence of foregrounds, we present the 2-D power spectra of the EoR signal (i.e. without foregrounds or thermal noise) in the 160--170 MHz band for the true values and the MCMC values -- the median values of $\zeta=21.13$, log$_{10}(T_{vir})=4.52$ , L$_X$/SFR = 40.00, and E$_0$=0.656 keV-- along with their ratio in Figure \ref{fig:ps_eor_params}. The true parameters give a maximum power that is up to 6 times higher than the MCMC parameters in the EoR window, suggesting that it is deceptively easy to discriminate between the models.

\begin{figure*}[]
\centering
\includegraphics[height= 0.3\textheight]{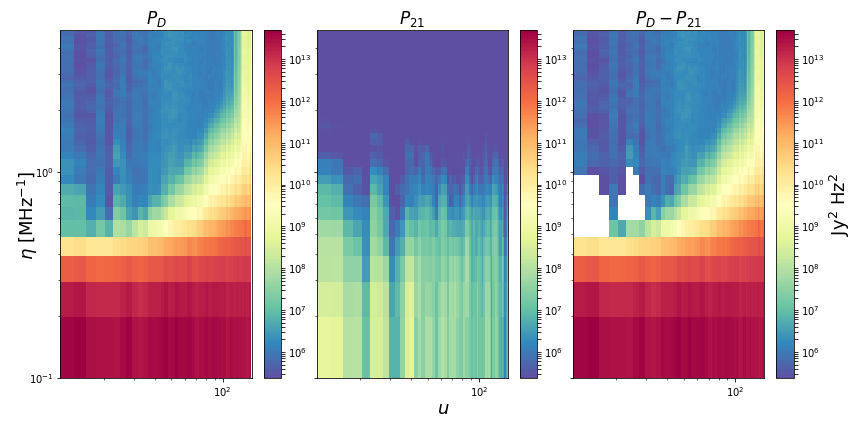}
\caption{\small\label{fig:ps_diff} The 2-D power spectra of the data i.e. EoR signal, foregrounds and thermal noise (left) and the EoR signal (middle) in the 160--170 MHz band along with their difference (right). The negative power (white region) in the right panel is caused by the unaccounted contribution of the cross-power terms between the EoR signal and the foregrounds and thermal noise.  }
\end{figure*}

In reality, however, it should be considered that the presence of foregrounds and noise may obscure modes that are vital for differentiation of these models. Figure \ref{fig:ps_diff} shows from left to right: the 2-D PS of the data i.e. the EoR signal, thermal noise and point source foregrounds, the 2-D PS of the true EoR signal and the difference between these two. The negative power in the right panel is caused by the unaccounted contribution of the cross-power terms between the EoR signal and the foregrounds and thermal noise. This is because the EoR signal and the expected thermal noise and foregrounds are modelled separately as mentioned in \S \ref{sec:bayesian_component}.

\begin{figure*}[]
\centering
\includegraphics[height= 0.5\textheight]{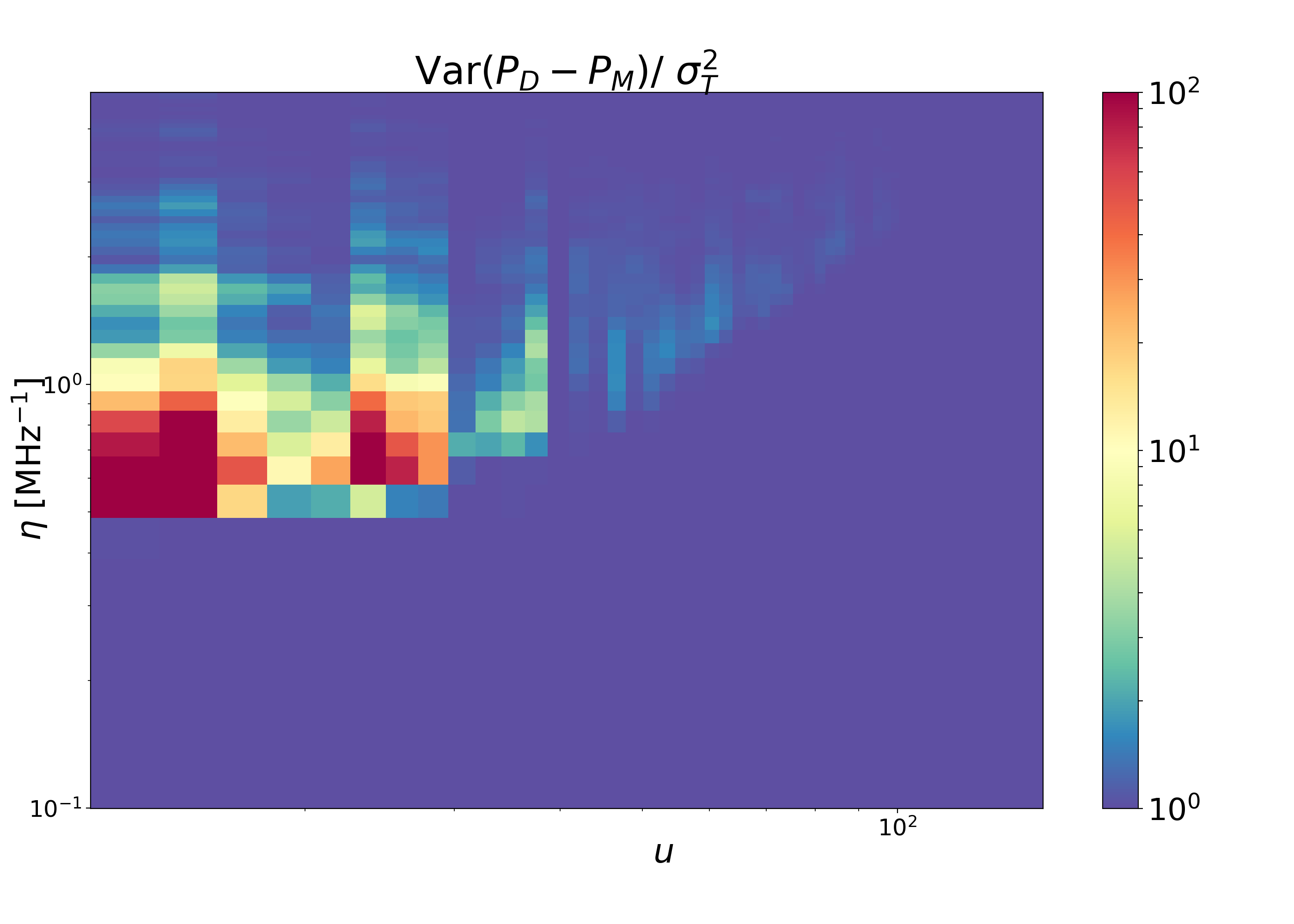}
\caption{\small\label{fig:var_residual} The ratio of the variance of the $P_D -P_M$ to the actual total variance, $\sigma_T$.}
\end{figure*}

We can estimate cross-power by looking at the variance of the difference between the 2-D power spectra of the data ($P_{D}$) and the model ($P_{M}$) for a single realization of 21cm power spectrum but with an ensemble of foregrounds and noise realizations following

\begin{align}
\label{eqn:var-residual}
\textrm{Var}(P_{D} - P_{M}) &= \rm Var [ \langle | (V_{21}+V_{FG}+V_{N})|^2 \rangle_{\textbf{u}} \nonumber \\
&- \langle |(V_{FG}+V_{N})|^2+ | (V_{21})|^2 \rangle_{\textbf{u}} ]  \nonumber \\
&= \langle 2 P_{21}[\textrm{Var}( V_{FG} )+\textrm{Var}( V_{N})] \rangle^*_{\textbf{u}},
\end{align}
where $V_{21}$, $V_{FG}$, and $V_{N}$ are the visibilities of the EoR signal, foregrounds and noise respectively. For simplification, $\langle$  $\rangle _{\textbf{u}}$ denotes the cylindrical average in \textbf{u} bin with weighting as described in \S \ref{sec:components_all} and $\langle$  $\rangle^* _{\textbf{u}}$ is the same as $\langle$  $\rangle _{\textbf{u}}$ except the weighting is squared; see the appendix for the complete derivation. From the equation, we can see that the variance is largest when $P_{21}$ is larger than the noise and foregrounds. To investigate this further, we present the ratio of the variance of $P_D -P_M$ and the true total variance $\sigma_T^2$ in Figure \ref{fig:var_residual}. The region that is more correlated than the other modes is at \textbf{u} $\leq 30$, whereby the size of the correlated bins are dictated by the correlation lengths of the window functions. This is evident from the red regions in the figure which extend over 2\textbf{u} by 4$\eta$ bins, corresponding to the correlation length of the MWA following $D/\lambda$ and the Blackman-Harris taper respectively.

\begin{figure*}[]
\centering
\includegraphics[height= 0.5\textheight]{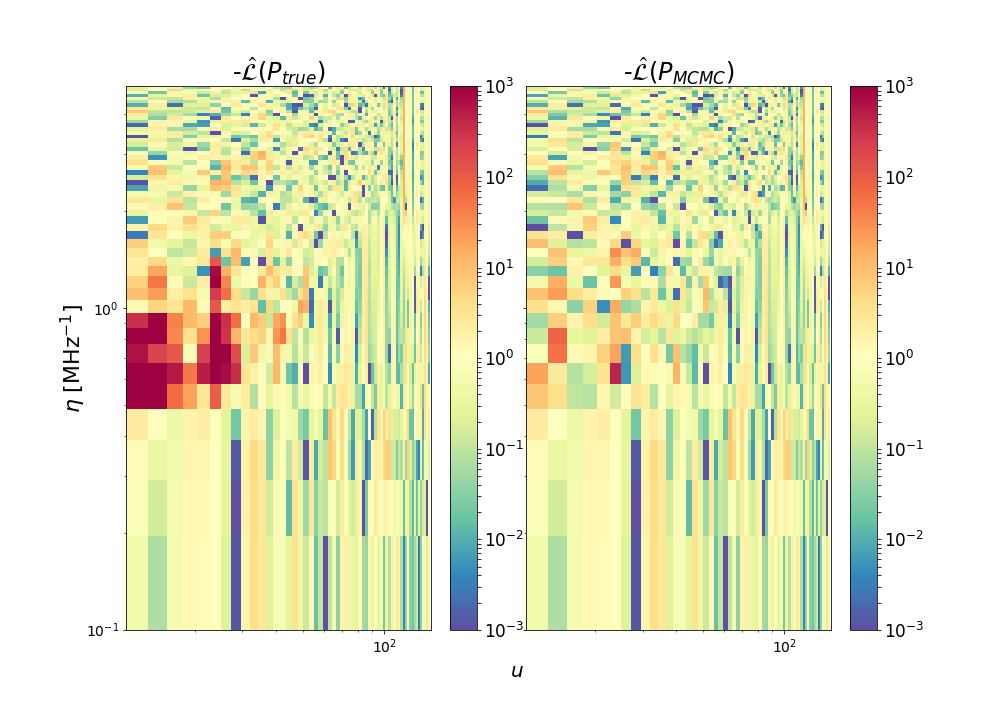}
\caption{\small\label{fig:pseudo_L} -$\mathcal{\hat{L}}$ for the true (left) and MCMC (right) parameters. -$\mathcal{\hat{L}}$ is overall greater for the true parameters, as apparent from the overwhelming red region in the left panel coinciding with low u.}
\end{figure*}

To further explore the region where most of the biasing takes place and gauge the net effect of the cross-power on the likelihood, we calculate the pseudo-likelihood ($\mathcal{\hat{L}}$) given by
\begin{align} 
\label{eqn:likelihood_pseudo}
\mathcal{\hat{L}} =  - \frac{(P_D - P_M)^2}{\sigma_T^2}.
\end{align}
This is a variation of Equation \ref{eqn:likelihood_new} but with $\sigma_T^2$ instead of $\Sigma$.  Figure \ref{fig:pseudo_L} show -$\mathcal{\hat{L}}$ for the true (left) and MCMC (right) parameters. -$\mathcal{\hat{L}}$ is overall greater for the true parameters, as apparent from the overwhelming red region in the left panel. This region coincides with low u, thus supporting our argument that this region is more correlated. With realistic datasets whereby the visibilities from the different components are summed before the PS is calculated, the likelihood has biased signal discrimination power when the same steps are not taken for the model. The likelihood favours whichever EoR model that accounts for the cross-power term thus resulting in biased constraints.

In principle, the signal to noise ratio is greater than one in every PS cell outside the wedge because of our low level of thermal noise. This results in the use of all modes of the PS in calculating the likelihood, including the ones that do not have any real constraining power, i.e. less signal. In addition, because the signal is stronger at modes with \textbf{u}$\leq 30$, these modes are more affected by the subtle cross-correlation effects from noise and foregrounds. The deficits in the modelling can thus lead to biased answers as the likelihood tries to compensate for the missing power.

Because our MCMC runs are shorter and may have not properly converged, the constraints from the MCMC may be invalid. To sample the ``true'' $1\sigma$ contours for a Gaussian posterior, we explore the $\Delta \chi^2$ contours by assuming that the true $\chi^2$ value is identical to the degree of freedom ($df$) (hence the $1\sigma$ level is $\sqrt{2 df}$).

\begin{figure*}[]
\centering
\includegraphics[height= 0.42\textheight]{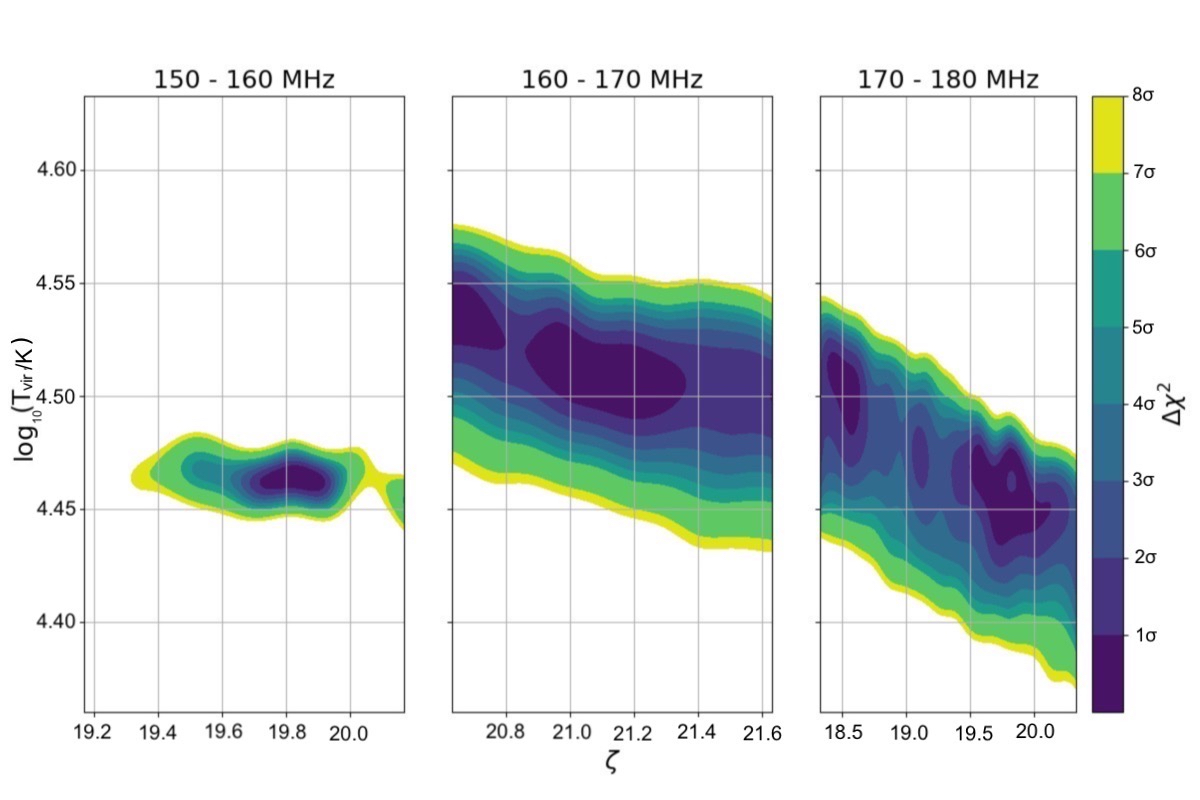}
\caption{\small\label{fig:chi} The $\Delta \chi^2$ contours for the three bands using the best fitting parameters from our framework which are presented in Table \ref{tbl:results}. With $df=2996,$ the $1\sigma$ level is $\sim 77$. Because the constraints for L$_X$/SFR and E$_0$ from the MCMC look reasonably inflated, we only focus on examining the $\Delta \chi^2$ contours for $\zeta$ and $T_{\rm vir}$. }
\end{figure*}

In Figure \ref{fig:chi}, we present the $\Delta \chi^2$ contours for the three bands using the best fitting parameters from the MCMC results shown in Figure \ref{fig:final_corner-3obs-noisefg}. With $df=2996,$ the $1\sigma$ level is $\sim 77$. Because the constraints for L$_X$/SFR and E$_0$ from the MCMC look reasonably inflated and unbiased, we focus on examining the $\Delta \chi^2$ contours for $\zeta$ and $T_{\rm vir}$.

We see that there are multiple regions in the parameter space that satisfy the $1\sigma$ condition in the middle and last frequency bands. This can confound the parameter estimation via MCMC. Additionally, we can see that the constraints for $T_{\rm vir}$ are not as tight as shown in Figure \ref{fig:final_corner-3obs-noisefg} but the constraints for the other parameters match the results from the MCMC.

We present the estimated median values of the parameters from each of the MCMC runs in Table \ref{tbl:results}, with the upper and lower limit being the 1$\sigma$ level from the $\Delta \chi^2$ grid. In the presence of both foregrounds and thermal noise, $T_{\rm vir}$, L$_X$/SFR and E$_0$ are all within 1$\sigma$ from the truth for the three frequency bands. For $\zeta$, however, the best fitting parameter values are off by as much as $5\sigma$ due to the unaccounted cross-power terms. This suggests that the effects of $\zeta$ on the PS space are primarily at low u region, since these modes are significantly impacted by the missing power.

\begin{table*}[]
\begin{tabular}{l|l|l|l|l|}
\cline{2-5}
                                         & \textbf{}     & \textbf{150 - 160 MHz}      & \textbf{160 - 170 MHz}       & \textbf{170 - 180 MHz}       \\ \hline
\multicolumn{1}{|l|}{\textbf{$\zeta$}}   & \textbf{20}   & $19.67^{+0.23}_{-0.00}$     & $21.13^{+0.18}_{-0.23}$      & $19.33^{+0.77}_{-0.00}$      \\ \hline
\multicolumn{1}{|l|}{\textbf{log$_{10}(T_{vir}$)}} & \textbf{4.48} & $4.47^{+0.01}_{-0.00}$      & $4.52^{+0.01}_{-0.04}$       & $4.46^{+0.04}_{-0.03}$       \\ \hline
\multicolumn{1}{|l|}{\textbf{log$_{10}(L_X/$SFR)}} & \textbf{40.5} & $40.09^{+0.48}_{-1.19}$     & $40.00^{+0.99}_{-0.92}$      & $40.22^{+0.94}_{-1.06}$      \\ \hline
\multicolumn{1}{|l|}{\textbf{E$_0$}}     & \textbf{500}  & $531.51^{+209.64}_{-34.39}$ & $656.50^{+409.39}_{-242.44}$ & $739.60^{+385.11}_{-348.29}$ \\ \hline
\end{tabular}
\caption{Results of astrophysical parameter constraints from the MCMC for the different frequency bands and with different components. The upper and lower limit show the 1$\sigma$ level from the $\Delta \chi^2$ grid.}
\label{tbl:results}
\end{table*}

\section{Discussion and Conclusion}
\label{sec:discussion}
Our \cmmc\ plug-in, \textsc{py21cmmc-fg}, simulates statistical point-source foregrounds and instrumental components that include a Gaussian beam, $uv$-sampling and thermal noise. It uses a Gaussian Fourier beam to re-grid the data before spherically averaging it to compute the 2-D PS. The 2-D PS allows for the use of the entire parameter space in the likelihood computation, effectively ensuring that the effects of foregrounds are accounted for, even in the EoR window.

We show that the implementation of a multivariate normal likelihood and the covariance of the noise and foregrounds appropriately down-weights modes that are contaminated by foregrounds, thus optimally accounting for all Gaussian information. Additionally, the low level of noise ensures that all modes in the EoR window are used in calculating the likelihood, even though not all modes actually have constraining power. Due to incomplete sampling of the posteriors and insufficient convergence of the chains, a cross-check with a Gaussian posterior distribution is needed to properly constrain the parameters. 

Furthermore, in order to accurately constrain the parameters, it is imperative that future experiments take into account the cross-power terms between the EoR signal and the foregrounds and noise. This can be done by either properly modelling all the components together, or by quantifying the cross-power terms separately and adding it to the framework we have developed. Nevertheless, both of these approaches may be computationally expensive and require a lot of memory because the gridded visibilities from all the foregrounds and thermal noise runs need to be stored and added to the EoR signal on each iteration.

Though our research does not address the full complexity of 21\,cm parameter estimation experiments, it is the first step towards a more realistic one. In the future, however, it would be useful to include: 
\begin{itemize}
    \item Earth rotation synthesis;
    \item the full curved-sky visibility equation, including the $w$-term particularly for the MWA;
    \item more realistic, non-Gaussian beams, including non-analytic ones such as the MWA beam \citep{sutinjo2015understanding};
    \item multi-redshift parameter estimation; 
    \item other systematics such as RFI, ionosphere, gain errors and cable reflections;
    \item diffuse Galactic foregrounds;
    \item clustered extragalactic point-source foregrounds; and
    \item uncertain foreground parameters.
\end{itemize}
Out of all these effects, the inclusion of Earth rotation synthesis and Galactic foregrounds are the most important because they can lead to a higher noise level up to 3$\times$ for longer baselines, as shown by \textsc{21cmSense}, and contain spatially-structured signal, for the latter.
Moreover, a further optimization of the code is necessary for more accurate constraints and absolute convergence of the MCMC.\\

\section*{Acknowledgements}

AN would like to thank R. J. J. Poulton and A. Bahramian for their contribution to this research. This work is supported by the Australian Research Council Centre of Excellence for All Sky Astrophysics in 3 Dimensions (ASTRO 3D), through project number CE170100013. CMT is supported by an ARC Future Fellowship through project number FT180100321. The International Centre for Radio Astronomy Research (ICRAR) is a Joint Venture of Curtin University and The University of Western Australia, funded by the Western Australian State government.
We acknowledge that the results in this paper have been achieved by using resources provided by the Pawsey Supercomputing Centre with funding from the Australian Government and the Government of Western Australia, and OzSTAR, funded by Swinburne and the Australian Governments Education Investment Fund.

\appendix
\section{Fourier Conventions}

The Fourier Transform (hereafter FT) conventions used in this paper are as follows. The continuous $n$-dimensional FT, $F(\textbf{k})_\infty$, can be written as 
\begin{equation}
\label{eqn:FT}
F(\textbf{k})_\infty = \sqrt[n]{\frac{|b|}{(2\pi)^{1-a}}} \int f(\textbf{r}) \exp(-ib\textbf{k} \cdot \textbf{r}) d^n\textbf{r},
\end{equation}
where \textbf{r} is the comoving coordinate in real space, \textbf{k} is the Fourier dual of \textbf{r}, and $a$ and $b$ are arbitrary constants \citep{murray2018powerbox}. Conventionally, radio interferometry uses $(a,b)=(0,2 \pi)$, while cosmology uses $(a,b)=(1,1)$. To approximate a continuous FT, a discrete FT normalises by the physical length of the data set ($L$) per number of data available for each dimension ($N$). The resulting modes, $k$, measured from the transform are
\begin{equation}
\label{eqn:FT_modes}
k= \frac{2 \pi}{b}\frac{m}{L},  m\in (-N/2,...,N/2).
\end{equation}\section{Variance of Residual}
\label{apdx:derivation}
Let $w_i$ be the weights used in averaging from 3D $(u,v,\eta)$-space down to 2D $(u,\eta)$-space, and $N^u_{uv}$ is the number of UV cells in a given $|u|$ bin; the variance of $P_D$ is thus
\begin{align*}
    \rm{Var}(P_D)&= \frac{\sum_{i \in uv}^{N^u_{uv}} w_i^2 \rm{Var}(\mathcal{Q})}{\left(\sum_{i \in uv}^{N^u_{uv}} w_i\right)^2} \\
    &\equiv  \langle \rm{Var}(\mathcal{Q}) \rangle^* _{\textbf{u}},
\end{align*}

where
\begin{equation}
    \mathcal{Q} = V_{21}V^{\dagger}_{21} + V_{FG}V^{\dagger}_{FG} + V_{N}V^{i\dagger}_{N} + 2\mathcal{R}e(V_{21} V^{\dagger}_{FG} + V_{21} V^{\dagger}_{N} + V_{FG} V^{\dagger}_{N}).
\end{equation}
In our case, while both $V_{\rm FG}$ and $V_N$ are complex random variables, $V_{21}$ is \textit{not} random. We use the same random seed to create the signal that comprises the mock data and all the model realizations throughout the MCMC (though these have different input astrophysics). Essentially, we treat the signal field as deterministic given the input parameters, and thus ignore cosmic variance. When taking the variance of $\mathcal{Q}$, this leads to all cross-terms involving ${\rm Var}(V_{21})$ or ${\rm Var}(P_{21})$ to be zero, leaving
\begin{align}
    {\rm Var}(\mathcal{Q}) = &{\rm Var}[P_{FG}] + {\rm Var}[P_{N}] \nonumber \\
    + &4{\rm Var}[\mathcal{R}e(V_{21}V^{\dagger}_{FG})]
    + 4{\rm Var}[\mathcal{R}e (V_{21}V^{\dagger}_{N})] 
    + 4{\rm Var}[\mathcal{R}e (V_{N}V^{\dagger}_{FG})] \nonumber \\
    + &4{\rm Cov}[V_{FG}V^{\dagger}_{FG}, \mathcal{R}e (V^{\dagger}_{21}V_{FG})]
    + 4{\rm Cov}[V_{N}V^{\dagger}_{N}, \mathcal{R}e (V^{\dagger}_{21}V_{FG})] \nonumber \\
    + &4{\rm Cov}[V_{N}V^{\dagger}_{N}, \mathcal{R}e (V^{\dagger}_{21}V_{N})]
    + 4{\rm Cov}[V_{N}V^{\dagger}_{N}, \mathcal{R}e (V^{\dagger}_{N}V_{FG})]\nonumber \\
    + &8{\rm Cov}[\mathcal{R}e (V_{21}V^{\dagger}_{N}), \mathcal{R}e (V^{\dagger}_{FG}V_{N})],
\end{align}
and the difference between the data and model variance is 
\begin{align}
    \Delta {\rm Var}(\mathcal{Q}) = &4{\rm Var}[\mathcal{R}e( V_{21}V^{\dagger}_{FG})]
    + 4{\rm Var}[\mathcal{R} e(V_{21}V^{\dagger}_{N})]  \nonumber \\
    + &4{\rm Cov}[V_{FG}V^{\dagger}_{FG}, \mathcal{R}e( V^{\dagger}_{21}V_{FG})]
    + 4{\rm Cov}[V_{N}V^{\dagger}_{N}, \mathcal{R}e( V^{\dagger}_{21}V_{FG})] \nonumber \\
    + &4{\rm Cov}[V_{N}V^{\dagger}_{N}, \mathcal{R}e( V^{\dagger}_{21}V_{N})] + 8{\rm Cov}[\mathcal{R}e( V_{21}V^{\dagger}_{N}), \mathcal{R}e( V^{\dagger}_{FG}V_{N})].
\end{align}
Assuming both $V_{FG}$ and $V_N$ to be proper complex random variables with a uniform phase distribution, we can simplify to
\begin{align}
    \Delta {\rm Var}(\mathcal{Q}) = &4 P_{21} {\rm Var}(V^\mathcal{R}_{FG})
    + 4 P_{21} {\rm Var}(V^\mathcal{R}_{N})  \nonumber \\
    + &4 (V^\mathcal{R}_{21} + V^\mathcal{I}_{21})  [\langle V^{3,\mathcal{R}}_{FG} \rangle - \langle V^{2,\mathcal{R}}_{FG} \rangle \langle V^\mathcal{R}_{FG} \rangle]
    + 4 (V^\mathcal{R}_{21} + V^\mathcal{I}_{21}) [\langle V^{3,\mathcal{R}}_{N} \rangle - \langle V^{2,\mathcal{R}}_{N} \rangle \langle V^\mathcal{R}_{N} \rangle] \nonumber \\
    + & 8 (V^\mathcal{R}_{21} + V^\mathcal{I}_{21}) \langle V^\mathcal{R}_{FG} \rangle {\rm Var}(V^\mathcal{R}_{N}),
\end{align}
where a superscript $\mathcal{R}$ indicates taking just the real component (which is statistically equivalent to the imaginary component).
Here the 3-point term in the noise disappears because it is symmetric about zero. 
Without explicitly modelling the 3-point term of the foregrounds, we also assume that it
is very much sub-dominant -- there is nothing to favour a negative over a positive foreground
visibility component. Recognising that ${\rm Var}(V^\mathcal{R}) = {\rm Var}(V)/2$, we finally obtain
\begin{align}
    \Delta {\rm Var}(\mathcal{Q}) = 2 P_{21} \left[{\rm Var}(V_{FG}) + {\rm Var}(V_{N}) \right].
\end{align}


\bibliography{bibliography}
%



\end{document}